%% file: main.tex
\documentclass[conference]{IEEEtran}

\usepackage{cite}
\usepackage{amsmath,amssymb,amsfonts}
\usepackage{algorithmic}
\usepackage{subcaption}
\usepackage[binary-units]{siunitx}
\usepackage{adjustbox}
\usepackage{svg}
\usepackage{graphicx}
\DeclareGraphicsExtensions{.pdf,.jpeg,.png,.svg}
\usepackage{caption}
\usepackage{textcomp}
\usepackage{xcolor}
\usepackage{url}
\usepackage{balance}

\usepackage{tikz}
\usetikzlibrary{shapes, calc}
\usetikzlibrary{positioning}
\usetikzlibrary{arrows.meta}
\usepackage{fontawesome5}

\def\BibTeX{{\rm B\kern-.05em{\sc i\kern-.025em b}\kern-.08em
    T\kern-.1667em\lower.7ex\hbox{E}\kern-.125emX}}
\begin{document}

\title{KuberneTSN: a Deterministic Overlay Network for Time-Sensitive Containerized Environments}

\author{
    \IEEEauthorblockN{Andrea Garbugli, Lorenzo Rosa, Armir Bujari, Luca Foschini}
    \IEEEauthorblockA{
        University of Bologna\\
        Department of Computer Science and Engineering\\
        Bologna, Italy \\
        name.surname@unibo.it}
}

\maketitle

\begin{abstract}
    The emerging paradigm of resource disaggregation enables the deployment of cloud-like services across a pool of physical and virtualized resources, interconnected using a network fabric. This design embodies several benefits in terms of resource efficiency and cost-effectiveness, service elasticity and adaptability, etc. Application domains benefiting from such a trend include cyber-physical systems (CPS), tactile internet, 5G networks and beyond, or mixed reality applications, all generally embodying heterogeneous Quality of Service (QoS) requirements. In this context, a key enabling factor to fully support those mixed-criticality scenarios will be the network and the system-level support for time-sensitive communication.
    Although a lot of work has been conducted on devising efficient orchestration and CPU scheduling strategies, the networking aspects of performance-critical components remain largely unstudied. Bridging this gap, we propose KuberneTSN, an original solution built on the Kubernetes platform, providing support for time-sensitive traffic to unmodified application binaries. We define an architecture for an accelerated and deterministic overlay network, which includes kernel-bypassing networking features as well as a novel userspace packet scheduler compliant with the Time-Sensitive Networking (TSN) standard. The solution is implemented as \textit{tsn-cni}, a Kubernetes network plugin that can coexist alongside popular alternatives. To assess the validity of the approach, we conduct an experimental analysis on a real distributed testbed, demonstrating that KuberneTSN enables applications to easily meet deterministic deadlines, provides the same guarantees of bare-metal deployments, and outperforms overlay networks built using the \textit{Flannel} plugin.
\end{abstract}

\begin{IEEEkeywords}
    time-sensitive networking, container, kubernetes, cloud continuum, network virtualization, bounded latency
\end{IEEEkeywords}

\vspace{-1em}

\section{Introduction}
\label{sec:intro}

The promise of edge computing is that of increasingly low latency, high bandwidth communication, and improved data security and privacy. Therefore, a stronger push for edge applications and service deployment is to be expected~\cite{bonomi}. However, in contrast to traditional cloud deployment environments, the edge has limited resources and may not be able to satisfy the overlapping and heterogeneous resource demands of all such applications. This fact has motivated researchers to extend the well-established cloud computing paradigm into the idea of \textit{edge-cloud computing} where an increasingly rich and heterogeneous set of resources between datacenters and the network edge, often called \textit{cloud continuum}, can be virtualized to host cloud-like services~\cite{Bittencourt2018}. The power of this paradigm relies on the combination of the well-known advantages of the cloud model, in particular flexibility, cost-effectiveness, and reconfigurability, with the performance advantage of running services as close to their final user as possible.

The success of this model is clear from its rapid and wide adoption in several heterogeneous domains, including application domains that embody time-sensitive requirements. As an example, the reference architecture of 5G and beyond standards relies on virtualized applications deployed in edge datacenters, or even co-located with the widely distributed base stations~\cite{trakadas-5G}. Control applications in the domains of Cyber-Physical Systems (CPS), Industrial Internet of Things, Tactile Internet, and in many other fields are increasingly pursuing the disaggregation trend, with virtualized application components deployed across the whole continuum of available resources, embodying heterogeneous Quality of Service (QoS) requirements, even among their internal components~\cite{tactilefitzek2021,ull-2}. Although many of those requirements can be easily met just by placing services physically closer to their final users, reducing key metrics such as latency or response time, core parts of these systems still struggle to balance strict performance demand with the overhead introduced by virtualization.

To mitigate this overhead, lightweight virtualization techniques like containerization have become the standard technology for platform-independent prototyping, development, and deployment of edge components. Compared to hypervisor-based virtual machines, containers are generally characterized by reduced overhead and higher scalability, representing a potential for innovation in service patterns, in virtue of setting up a unified service provisioning platform capable of adhering to applications' QoS specifications~\cite{containervm}.

Furthermore, containers are seamlessly integrated into resource management and orchestration platforms, with Kubernetes in its full or reduced versions (e.g., k3s) as the \textit{de-facto} standard technology~\cite{kubernetes}. Resource management and orchestration are paramount in the edge cloud, as it automatically deploys, monitors, and migrates containerized application components across the shared infrastructure, enforcing applications' QoS specifications.

However, containerization alone is not a panacea. Given the highly distributed nature of edge cloud applications, specific attention to network and system-level aspects is paramount to effectively support the most performance-demanding components. Yet, previous work mostly focused on efficient orchestration and CPU scheduling of containers~\cite{toka2021ultra,k8-latency-orch,sensors-tsn}, leaving those aspects largely unstudied.

In this paper, we design a cost-efficient solution to enable \textit{accelerated and deterministic communication} among containerized applications. To this end, we define a novel architecture for a container overlay network that combines two techniques for high-performance communication. First, we adopt a form of kernel-bypassing networking to remove the overhead of the kernel networking stack~\cite{understanding}. Second, we propose a novel userspace packet scheduler, compliant with the Time-Sensitive Networking (TSN) standard, to allow the time-bounded data distribution and communication among networked components~\cite{tsn}. We implement our proposal as \textit{tsn-cni}, a novel Kubernetes network plugin that can be seamlessly integrated alongside existing options (e.g., Flannel, Calico). This way, application designers are free to choose the most appropriate support for traffic flows with different degrees of criticality. Finally, we evaluate \textit{tsn-cni} on a real testbed, showing that containerized TSN applications can achieve determinism and performance comparable to bare metal applications, and better than using the network fabric set up by the popular \textit{Flannel} plugin.

\section{Background}
\label{sec:background}

This section provides a brief introduction to container overlay networks, their rationale, and support in the Kubernetes platform. Next, we provide a concise background on the TSN standard and kernel-bypassing techniques.

\subsection{Container Overlay Networks}

\begin{figure}[t]
    \centering
    \includegraphics[width=.7\columnwidth]{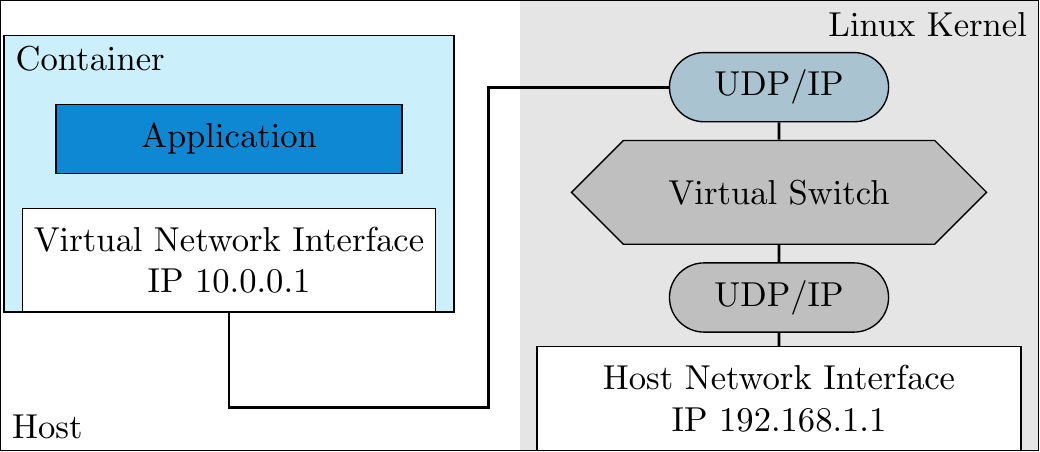}
    \caption{Container networking in overlay mode. \vspace{-1em}}
    \label{fig:containernetworking}
\end{figure}

Containers generally have four networking modes available: bridge, host, macvlan, and overlay. The overlay mode is the most popular, especially in combination with Kubernetes, as it provides better isolation, ease of use, and security; hence we limit our description to this scheme. In this mode, as depicted in Fig.~\ref{fig:containernetworking}, containers are connected on an overlay network, potentially spanning multiple physical nodes even across different networks. On each container, a virtual network interface is created, to which applications can assign an arbitrary IP address.
This interface is connected to the outside through a virtual switch, located in the host operating system kernel, which has two main roles: it works as a network bridge to allow communication among co-located containers, and it tunnels network traffic toward the remote container(s) across the physical network. This way, containers on the same overlay network have an isolated address namespace and configuration settings, disjoint from the host network or from other overlays.

When using Kubernetes, by default each container has a single network interface for all the network traffic, including management and control plane interactions (e.g., with the Kubernetes master). To distinguish among different traffic classes, the Multus plugin~\cite{multus} allows attaching additional interfaces to containers. Multus is a meta-plugin, as it defines a \textit{container network interface} (CNI) that other plugins can implement to configure a Layer 3 network fabric and optionally provide additional advanced features. Several such plugins are available, such as Flannel, Calico, or Weave. Unfortunately, none of those supports the definition of an accelerated and deterministic communication channel among containers. Compared to these alternatives, in this work we design a novel plugin architecture to offer such guarantees. We still rely on a virtual switch, but we move the sender-side datapath to userspace and provide a novel packet scheduler compliant with the TSN standard. This choice allows users to obtain enhanced network performance with no modifications to application binaries and atop off-the-shelf hardware and operating systems, without requiring any patches or specific configurations from the final user.

\subsection{Time-Sensitive Networking}
Designed to support soft real-time industrial traffic, the set of standards grouped under the name of Time-Sensitive Networking aims to introduce determinism to IEEE 802.1 networks via a set of features, including but not limited to time synchronization, programmability, etc.~\cite{tsn}. First, TSN requires that all the communication participants share a unique time reference, and the IEEE 802.1AS standalone protocol provides an adequate mechanism to ensure this synchronization~\cite{nasrallah2019}. A second key concept in TSN is packet scheduling. The IEEE 802.1Qbv standard defines a traffic shaper, called Time-Aware Shaper (TAS), that can prioritize the frames belonging to classes of traffic with different time criticality. This prioritization is based on time-aware communication windows, called \emph{time-aware traffic windows}, that repeat cyclically. Each window is divided into \emph{time slots} that can be associated to different traffic classes: frames belonging to the same class are buffered until the next opening of the associated time slot. This way, TSN guarantees bounded latency and jitter to time-critical traffic, as well as no interference from best-effort traffic.

From a practical viewpoint, to enable this kind of communication, developers must configure the kernel-based Traffic Controller (TC), which implements a TAS shaper, to set up the desired number of traffic classes, their priority, and time slots duration. Then, applications open a datagram socket with the \texttt{SO\_TXTIME} flag, so that they can associate a desired transmission time to each outgoing message. Unfortunately, there are two obstacles to the adoption of this standard from containerized environments. First, we noted that some OS images do not support the \texttt{SO\_TXTIME}. Second, the transmission time is never forwarded outside the container network namespace to the virtual switch. To overcome these limitations, KuberneTSN intercepts the container TSN traffic and forwards it to a novel userspace scheduler, responsible to enforce the TAS shaping. This component, which replaces the Linux-specific kernel-based scheduler, is the key architectural element that we leverage to provide time-sensitive networking features to containerized applications, and it is fully integrated into the \textit{tsn-cni} Kubernetes plugin.

\subsection{Kernel-bypassing Networking}

In a container overlay network, each outgoing packet must cross the networking stack twice, one in its isolated network namespace and one in the host namespace, and must also cross through a virtual switch (Fig.~\ref{fig:containernetworking}). The combination of all these steps adds significant per-packet communication overhead~\cite{slim-nsdi19}, unacceptable for time-sensitive edge applications.

In recent years, several \textit{kernel-bypassing} networking approaches, also known as network acceleration techniques, have emerged to support performance-critical applications. Among them, the Data Plane Development Kit (DPDK)~\cite{dpdk} is an increasingly popular library that adopts this approach without requiring special hardware or OS support. DPDK lets applications access a userspace version of the network device drivers (\textit{Poll Mode Drivers}) to directly send or receive Ethernet packets on the network. Applications and drivers exchange data through a shared memory area registered with the network card for Direct Memory Access (DMA), thus communication is \textit{zero-copy} and avoids kernel/user context changes. This way, communication is much more efficient, and, in principle, applications in the edge cloud would immensely benefit from the related performance improvements. However, DPDK exposes a low-level C interface, very difficult to use and scarcely integrated within virtualization engines~\cite{insane}.

In KuberneTSN, we accelerate the outgoing container data path using DPDK transparently to user applications. Specifically, we design KuberneTSN to bypass the kernel networking stack in the container namespace, sending data directly to a userspace virtual switch. Then, we adopt a userspace version of a widely used and open-source virtual switch, Open vSwitch (OVS)~\cite{ovs}, which in turn uses DPDK to bypass the kernel networking stack in the host namespace.

Overall, KuberneTSN combines three well-known networking approaches, namely overlay networks, TSN scheduling, and kernel-bypassing networking, and leverages them to offer the option of a deterministic and accelerated inter-container communication, well integrated into the state-of-the-art Kubernetes orchestrator and complementary to existing networking approaches for best-effort traffic.

\section{Related Work}
\label{sec:related}

Previous research on the containerization of critical application components mainly focused on orchestration strategies and CPU scheduling~\cite{sensors-tsn,toka2021ultra}. These works investigate the best strategies to place components on suitable resources and ensure that those resources can schedule the execution of containerized applications according to their requirements. Yet, they never take network and system-related aspects into account. We consider these works complementary to our proposal, as we envision that network and computing resources for edge applications should be orchestrated together.

Despite the importance of networking for edge applications, researchers paid less attention to the networking requirements of critical applications. Abeni et al.~\cite{Abeni2020.2} evaluate different kernel-bypass approaches for inter-container communications, outlining the great potential of DPDK as network accelerator compared to the kernel-based approach. However, their contribution is limited to a framework for performance evaluation.

Slim~\cite{slim-nsdi19} proposes a solution to reduce the processing overhead on container overlay networks. At its crux, the proposal avoids processing packets multiple times on the same host (see Sec. \ref{sec:background}); instead, it defines a component that intercepts calls to the socket API and directly translates network addresses from the overlay into the host namespace (and vice versa). This way, packets traverse the kernel networking stack only once. SocksDirect~\cite{socksdirect} uses the same interception technique to re-route packets on an accelerated kernel-bypassing datapath, but this is possible only with the \textit{host} container networking mode. Both these works introduce the idea of accelerating container inter-networking, and both show significant performance advantages for a wide range of applications built on top of them. However, these solutions are not integrated with standard production-ready technologies such as Kubernetes. Furthermore, as they target datacenter environments, their focus is on accelerated support for reliable connection-oriented transport protocols (TCP), and they do not provide any support for time-sensitive applications such as TSN, a key requirement for edge applications. In this work, we adopt similar techniques (socket interception, kernel-bypassing) to accelerate network operations, but we also provide guarantees on connection determinism (through TSN) and implement our solution as a plugin for highly standard development and deployment technologies.

Finally, the use of TSN in virtual environments is a relatively new trend, as the standard was originally intended for bare-metal industrial applications. Leonardi et al.~\cite{Leonardi2020} first hypothesized this possibility, identifying three distinct architectural approaches to enhance hypervisor-based virtualization with time-triggered communication. In a previous work~\cite{tsn-vm}, we showed for the first time on a real testbed that TSN applications can execute in remote virtual machines, embodying even better performance than bare-metal thanks to the adoption of kernel-bypassing techniques. In this paper, we target containerized applications and take a step further by implementing our solution as a Kubernetes network plugin, thus allowing an application to select the most appropriate overlay network meeting their requirements.

\section{KuberneTSN: an Accelerated and Deterministic Overlay Network}
\label{sec:detoverlay}

\begin{figure}[t]
    \centering
    \includegraphics[width=\columnwidth]{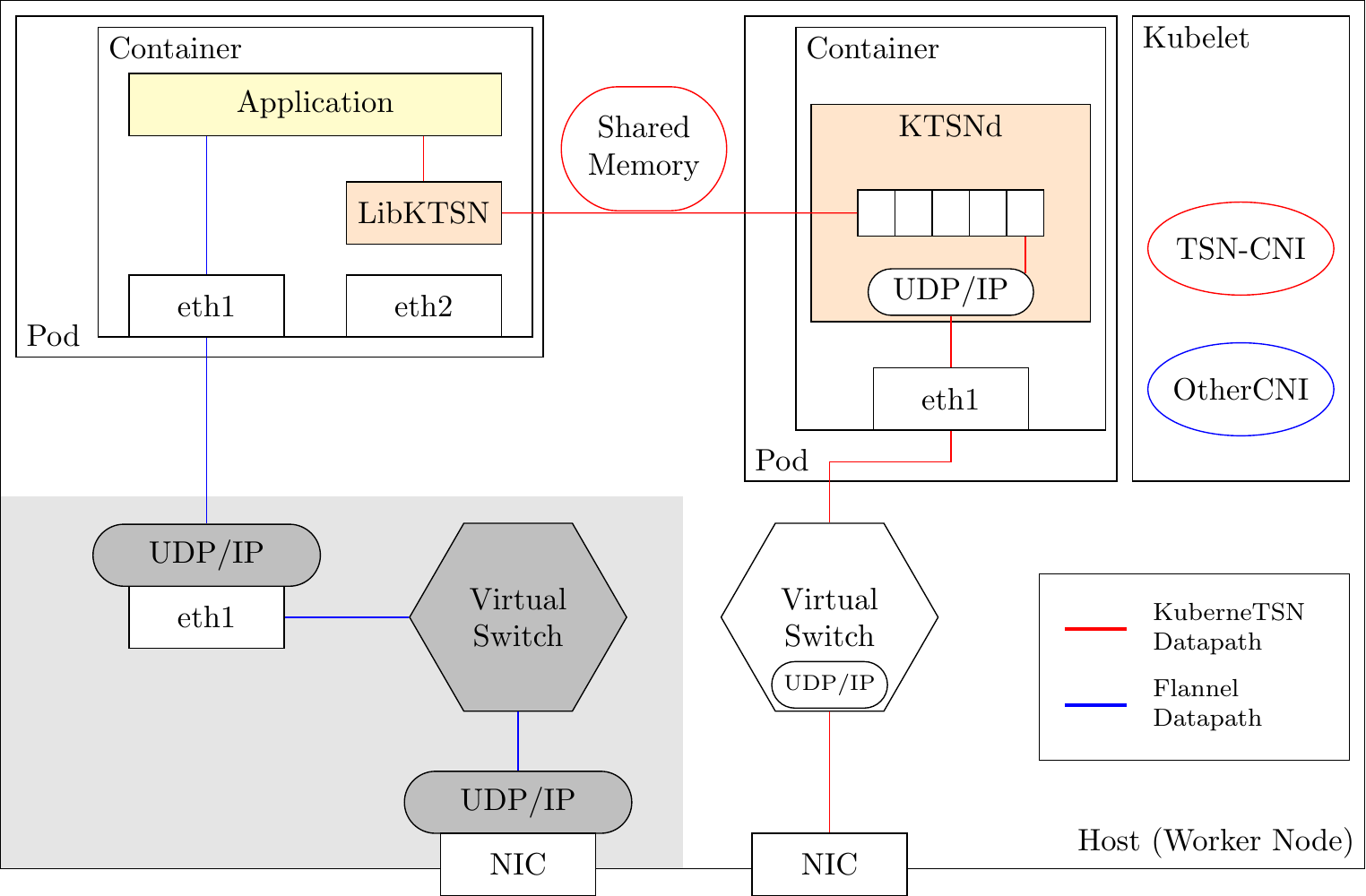}
    \caption{The architecture for an accelerated and deterministic overlay network, implemented as a Kubernetes CNI plugin.}
    \label{fig:architecture}
\end{figure}

KuberneTSN defines the architecture for a novel \textit{accelerated} and \textit{deterministic} container overlay network, addressing the time-sensitive requirements of containerized business or control logic. To achieve this goal, we intervene and modify the packet processing pipeline for the \textit{outgoing} container traffic through the use of two novel architectural components: a user library named \textit{LibKTSN}, and a daemon named \textit{KTSNd}. Fig.~\ref{fig:architecture} shows those components and the role they play in the definition of a new data path for time-sensitive traffic.

\textit{LibKTSN} exposes the standard POSIX socket interface to the application binaries. This way, any time the application issues a send operation on a datagram socket, the library intercepts it and forwards the packets to a memory area shared with the KTSNd daemon. We are interested in servicing time-sensitive traffic, so we only capture outgoing transmissions that have an explicit transmission time, i.e., TSN traffic, with the \texttt{SO\_TXTIME} socket option. Otherwise, packets are forwarded onto the regular data path. This approach enables TSN networking regardless of the container images, unlike the currently available alternatives (see Sec.~\ref{sec:background}). LibKTSN is the only component of our solution that should be present in the application container. We provide it as a shared library and use the flag \texttt{LD\_PRELOAD} to transparently intercept traffic: hence, no changes are required to the application code.

The \textit{KTSNd} daemon represents the key component of our proposal, as it works both as a packet scheduler and a network accelerator. Once it detects a new packet from an application, KTSNd schedules its actual transmission based on the application-provided transmission time. Although we design the daemon to be agnostic to the specific scheduling strategy, by default it works as a Time-Aware Shaper (TAS) compliant with the IEEE 802.1Qbv standard (see Sec.~\ref{sec:background}). Currently, this packet scheduling option is not available for containerized applications, as popular virtual switches (e.g., Linux bridge, Open vSwitch, etc.) do not support it. Therefore, our solution is the first to provide deterministic packet scheduling for unmodified application binaries running in containers.

When time comes to transmit a scheduled packet, the scheduler must send it on the network on behalf of the original application, preserving the source MAC, IP addresses, and UDP ports, and minimizing the packet processing delays to meet the user-required transmission time as precisely as possible. To satisfy these requirements, we adopt a kernel-bypassing approach and move the entire transmission pipeline in userspace. This way, we avoid the expensive double-crossing of the kernel networking stack and the unnecessary user/kernel thread context switches (see Sec.~\ref{sec:background}) and instead provide our own simple and efficient implementation of the UDP/IP stack directly within KTSNd, using the DPDK library to forward packets on the virtual L2 link. This choice allows us to preserve the original packet metadata, as we can manipulate protocol headers directly, and significantly reduce the processing overhead. As shown in Fig.~\ref{fig:architecture}, packets are then handled by a userspace virtual switch that, in turn, should provide its own UDP/IP userspace stack to forward them on the physical network. In our implementation, we adopt a widely-used, state-of-the-art userspace virtual switch, Open vSwitch~\cite{ovs}, which also uses DPDK for kernel-bypassing.

The simple yet powerful design makes KuberneTSN easy to integrate into standard platforms such as the Kubernetes orchestrator in its various distributions, making it ready to use for critical networked applications embodying stringent requirements. To this aim, we build a Kubernetes network plugin, \textit{tsn-cni}, that implements our architecture. Specifically, \textit{tsn-cni} implements the Multus CNI interface~\cite{multus} and thus a Layer 3 network fabric that includes our accelerated and deterministic data path. The plugin requires applications to include LibKTSN in their execution environment, and it encapsulates the KTSNd daemon in a separate container. This approach is strategic to support time-sensitive edge applications: because multiple network plugins can be used at the same time, developers can choose standard ones (e.g., Flannel, Calico) for best-effort traffic, and \textit{tsn-cni} for time-sensitive networking, as represented in Fig.~\ref{fig:architecture}. Therefore, KuberneTSN and its \textit{tsn-cni} implementation enhance the capabilities of the edge-cloud not only by supporting deterministic networking but also by integrating this option in a familiar ecosystem for application designers. By tagging application components as time-sensitive, they can instruct Kubernetes to automatically deploy KTSNd alongside the application containers, thus transparently obtaining support for performance-sensitive workloads.

\begin{figure*}[ht]
    \begin{subfigure}[t]{0.5\textwidth}
        \centering
        \resizebox{.95\linewidth}{!}{\input{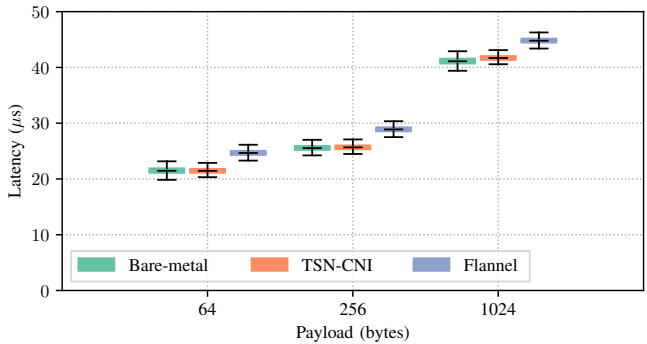}
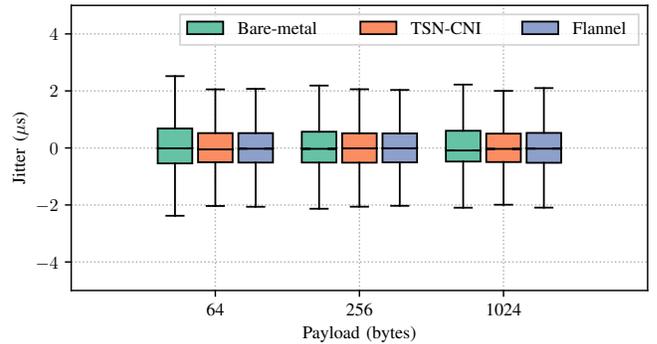}
        \caption{Latency.}
        \label{subfig:latency}
    \end{subfigure}%
    \begin{subfigure}[t]{0.5\textwidth}
        \centering
        \resizebox{.95\linewidth}{!}{\input{jitter.pgf}}
        \caption{Jitter.}
        \label{subfig:jitter}
    \end{subfigure}
    \caption{Performance comparison among three deployment options for the latency test application: bare metal, containerized with \textit{tsn-cni}, containerized with \textit{Flannel}. The experiment is repeated for increasing payload sizes: \SI{64}{\byte}, \SI{256}{\byte}, \SI{1024}{\byte}.\vspace{-1em}}
    \label{fig:comparison}
\end{figure*}

\section{Experimental evaluation}
\label{sec:quantitative evaluation}

In this section, we evaluate the performance of the \textit{tsn-cni} plugin, which implements the KuberneTSN architecture. The purpose of the experimental assessment is twofold: on the one hand, we want to show that the \textit{accelerated} datapath we propose is indeed faster than the current state-of-the-art networking options; on the other hand, we demonstrate that our solution can in fact provide \textit{deterministic} guarantees. In particular, we compare \textit{tsn-cni} against two alternatives. The first is a bare metal setting that reproduces the way typical TSN applications are deployed, in order to assess the overhead introduced by the virtualization layer. The second is \textit{Flannel}, a popular CNI plugin for Kubernetes. In its recommended configuration, Flannel uses a Linux bridge in combination with VXLAN encapsulation to implement the virtual switch, thus building an overlay network that corresponds to the \textit{regular datapath} of Fig.~\ref{fig:architecture}. By comparing \textit{tsn-cni} and Flannel, we assess whether KuberneTSN meets its design goal of providing additional performance benefits and deterministic properties to inter-container networking.

For the purpose of this evaluation, we build a simple TSN application consisting of two processes, a talker and a listener, each running inside a container on two remote hosts. We then set up a latency test in which the talker sends UDP packets with a cycle of \SI{1}{\milli\second}. The test measures two representative indicators of time-sensitive communications: end-to-end latency and jitter. The end-to-end latency of a message is defined as the time interval between the time of transmission predicted by the talker, sometimes also called transmission time, and the time of actual reception by the listener. The jitter measures how much the actual arrival time of each message differs from the expected arrival time: more precisely, if $t_i$ is the arrival time of the $i$-th message, its jitter is defined as $Jitter(i) = t_i - (t_{i-1} + T)$, where $T$ is the transmission period (in this work, $T=1ms$). It is noteworthy to point out that the bare-metal and the \textit{tsn-cni} test suites are implemented as actual TSN applications, which associate a desired transmission time to each packet. However, for the test adopting Flannel as a communication choice, this option is not available, as the TSN scheduling would not be enforced (see Sec.~\ref{sec:background}). Instead, the only alternative is to send one message and then sleep, repeating this behavior every $T$.

\subsection{Experimental Settings}
The evaluation analysis is conducted on a real testbed which reproduces an edge deployment scenario. The testbed comprises two Dell Workstations, each equipped with an Intel I225 NIC, an Intel i9-10980XE 18/36 CPU, and \SI{64}{\giga\byte} RAM. The two hosts are interconnected through a physical TSN-compliant switch. Each host runs Ubuntu 22.04 with Linux kernel 5.16. When using Open vSwitch~\cite{ovs}, we adopt its two variants, the kernel-bypassing on the sender side, and the kernel-based on the receiver side. As required by TSN, the clocks of the two hosts are synchronized using two PTP daemons. Finally, we pin the processes to dedicated cores so to avoid any bias in the measurements induced by the CPU scheduling policy.

\subsection{End-to-end Latency}

Figure~\ref{subfig:latency} reports the end-to-end latency and jitter measured for three typical data sizes (\SI{64}{\byte}, \SI{256}{\byte}, \SI{1024}{\byte}) for each of the considered deployment scenarios: bare-metal and containerized applications with \textit{tsn-cni} or Flannel as network plugin. A first consideration is that the performance of \textit{tsn-cni} is always very good, with median latency values ranging from \SI{21.5}{\micro\second} in the case of small packets (\SI{64}{\byte}) to \SI{41.7}{\micro\second} for \SI{1024}{\byte}. These values are almost identical to those registered for the bare metal deployment, with a small variation in the \SI{}{\nano\second} scale starting to appear for the 1KB packet size. Latency variability is negligible in both cases.
If we consider Flannel, we note a slight, but evident latency increase (\SI{12}{\percent} on average). This is the result of the expensive in-kernel packet processing, which we avoid thanks to the kernel bypassing technique embodied in our solution. The same trend observed for latency is confirmed by the analysis of the jitter metric reported in Fig.~\ref{subfig:jitter}: the median value is zero in almost all cases and the variability is negligible. Therefore, we can conclude that KuberneTSN and its \textit{tsn-cni} implementation succeed in minimizing the packet processing overhead for containerized applications, achieving the goal of an \textit{accelerated} data path.

Overall, our experiments show that both \textit{tsn-cni} and Flannel show good latency numbers, although our kernel-bypassing solution shows lower median values. In principle, one could expect even better performance from \textit{tsn-cni}, as raw DPDK is particularly fast~\cite{Abeni2020.2}. However, we noted that the OVS-DPDK implementation introduces a non negligible overhead on our userspace datapath, consisting of at least 23\% of the total reported latency. Nevertheless, we decided to keep it in our system as it is a widely-used tool, supported by an active community. Even more importantly, while still demonstrating better performance, it supports a rich set of additional features for virtual networking, e.g. OpenFlow programmability, compared to the basic Linux bridges used by Flannel.

\subsection{Determinism}

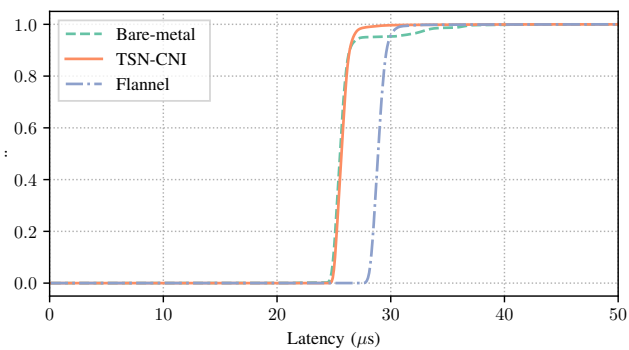
\begin{figure}
    \begin{center}
        \adjustbox{max width=0.99\columnwidth} {
            \input{cdf.pgf}
        }
    \end{center}
    \caption{CDF with packets of 256 bytes.}
    \label{fig:cdf}
\end{figure}

To assess whether KuberneTSN can effectively provide deterministic guarantees to time-sensitive flows, we consider again the latency test results discussed before, but in Fig.~\ref{fig:cdf} we plot the respective Cumulative Distribution Function (CDF). Ideally, the curve should be as vertical as possible, implying a highly predictable packet reception time. In this context, the bare metal application and the containerized application using \textit{tsn-cni} show overlapping performance, very close to the ideal behavior. In particular, for \textit{tsn-cni} the \SI{90}{\percent} and the \SI{99}{\percent} probability correspond to \SI{26.4}{\micro\second} and \SI{28.1}{\micro\second} respectively. Instead, for Flannel these thresholds are \SI{29.6}{\micro\second} and \SI{30.7}{\micro\second} respectively, implying a less precise arrival time interval.

This difference demonstrates the advantage of using KuberneTSN for time-sensitive traffic. The main reason for this behavior is the way the test application sends messages: when using Flannel, we cannot explicitly set a transmission time, as this feature is not supported in current containerized environments. Hence, we are constrained to fall back to a classic send-and-sleep loop, mimicking a periodic send operation. The effect of this difference is minimal in our experiment, as we do not have other competing flows; however, previous work~\cite{tempos} demonstrates that time-sensitive flows require dedicated support. \textit{tsn-cni} serves this purpose by providing essential support to containerized applications so as to meet heterogeneous flow requirements in mixed-criticality scenarios.

\section{Conclusion and future work}
\label{sec:conclusion}

We presented KuberneTSN, an architecture for an accelerated and deterministic container overlay network. KuberneTSN defines a novel userspace TSN packet scheduler and adopts a kernel-bypassing approach to minimize packet processing delays. We implemented KuberneTSN as a network plugin for the Kubernetes orchestrator, called \textit{tsn-cni}, so that it can be used alongside existing network fabrics to better support time-sensitive edge applications. The solution was evaluated on a real testbed, showing that containerized applications using \textit{tsn-cni} have the same level of performance and determinism as bare metal applications, outperforming the widely used Flannel network plugin.

Future work will include a detailed performance characterization of KuberneTSN under different traffic conditions, and a demonstration of the use of \textit{tsn-cni} in combination with other network plugins. In the longer term, as performance-demanding AI/ML components are increasingly moved to the network edge, we are interested in a systematic performance study of the inter-container datapath to highlight further optimization opportunities.

\section*{Acknowledgements}
This work was partially supported by the H2020 TERMINET project (Grant agreement \#: 957406).

\bibliographystyle{IEEEtran}
\bibliography{IEEEabrv,refs}
\balance
\end{document}

%% file: jitter.pgf
%% Creator: Matplotlib, PGF backend
%%
%% To include the figure in your LaTeX document, write
%%   \input{<filename>.pgf}
%%
%% Make sure the required packages are loaded in your preamble
%%   \usepackage{pgf}
%%
%% Also ensure that all the required font packages are loaded; for instance,
%% the lmodern package is sometimes necessary when using math font.
%%   \usepackage{lmodern}
%%
%% Figures using additional raster images can only be included by \input if
%% they are in the same directory as the main LaTeX file. For loading figures
%% from other directories you can use the `import` package
%%   \usepackage{import}
%%
%% and then include the figures with
%%   \import{<path to file>}{<filename>.pgf}
%%
%% Matplotlib used the following preamble
%%   
%%   \makeatletter\@ifpackageloaded{underscore}{}{\usepackage[strings]{underscore}}\makeatother
%%
\begingroup%
\makeatletter%
\begin{pgfpicture}%
\pgfpathrectangle{\pgfpointorigin}{\pgfqpoint{5.000000in}{2.700000in}}%
\pgfusepath{use as bounding box, clip}%
\begin{pgfscope}%
\pgfsetbuttcap%
\pgfsetmiterjoin%
\definecolor{currentfill}{rgb}{1.000000,1.000000,1.000000}%
\pgfsetfillcolor{currentfill}%
\pgfsetlinewidth{0.000000pt}%
\definecolor{currentstroke}{rgb}{1.000000,1.000000,1.000000}%
\pgfsetstrokecolor{currentstroke}%
\pgfsetdash{}{0pt}%
\pgfpathmoveto{\pgfqpoint{0.000000in}{0.000000in}}%
\pgfpathlineto{\pgfqpoint{5.000000in}{0.000000in}}%
\pgfpathlineto{\pgfqpoint{5.000000in}{2.700000in}}%
\pgfpathlineto{\pgfqpoint{0.000000in}{2.700000in}}%
\pgfpathlineto{\pgfqpoint{0.000000in}{0.000000in}}%
\pgfpathclose%
\pgfusepath{fill}%
\end{pgfscope}%
\begin{pgfscope}%
\pgfsetbuttcap%
\pgfsetmiterjoin%
\definecolor{currentfill}{rgb}{1.000000,1.000000,1.000000}%
\pgfsetfillcolor{currentfill}%
\pgfsetlinewidth{0.000000pt}%
\definecolor{currentstroke}{rgb}{0.000000,0.000000,0.000000}%
\pgfsetstrokecolor{currentstroke}%
\pgfsetstrokeopacity{0.000000}%
\pgfsetdash{}{0pt}%
\pgfpathmoveto{\pgfqpoint{0.510806in}{0.456793in}}%
\pgfpathlineto{\pgfqpoint{4.958330in}{0.456793in}}%
\pgfpathlineto{\pgfqpoint{4.958330in}{2.658330in}}%
\pgfpathlineto{\pgfqpoint{0.510806in}{2.658330in}}%
\pgfpathlineto{\pgfqpoint{0.510806in}{0.456793in}}%
\pgfpathclose%
\pgfusepath{fill}%
\end{pgfscope}%
\begin{pgfscope}%
\pgfpathrectangle{\pgfqpoint{0.510806in}{0.456793in}}{\pgfqpoint{4.447524in}{2.201537in}}%
\pgfusepath{clip}%
\pgfsetbuttcap%
\pgfsetroundjoin%
\pgfsetlinewidth{0.803000pt}%
\definecolor{currentstroke}{rgb}{0.690196,0.690196,0.690196}%
\pgfsetstrokecolor{currentstroke}%
\pgfsetdash{{0.800000pt}{1.320000pt}}{0.000000pt}%
\pgfpathmoveto{\pgfqpoint{1.622687in}{0.456793in}}%
\pgfpathlineto{\pgfqpoint{1.622687in}{2.658330in}}%
\pgfusepath{stroke}%
\end{pgfscope}%
\begin{pgfscope}%
\pgfsetbuttcap%
\pgfsetroundjoin%
\definecolor{currentfill}{rgb}{0.000000,0.000000,0.000000}%
\pgfsetfillcolor{currentfill}%
\pgfsetlinewidth{0.803000pt}%
\definecolor{currentstroke}{rgb}{0.000000,0.000000,0.000000}%
\pgfsetstrokecolor{currentstroke}%
\pgfsetdash{}{0pt}%
\pgfsys@defobject{currentmarker}{\pgfqpoint{0.000000in}{-0.048611in}}{\pgfqpoint{0.000000in}{0.000000in}}{%
\pgfpathmoveto{\pgfqpoint{0.000000in}{0.000000in}}%
\pgfpathlineto{\pgfqpoint{0.000000in}{-0.048611in}}%
\pgfusepath{stroke,fill}%
}%
\begin{pgfscope}%
\pgfsys@transformshift{1.622687in}{0.456793in}%
\pgfsys@useobject{currentmarker}{}%
\end{pgfscope}%
\end{pgfscope}%
\begin{pgfscope}%
\definecolor{textcolor}{rgb}{0.000000,0.000000,0.000000}%
\pgfsetstrokecolor{textcolor}%
\pgfsetfillcolor{textcolor}%
\pgftext[x=1.622687in,y=0.359571in,,top]{\color{textcolor}\rmfamily\fontsize{10.000000}{12.000000}\selectfont 64}%
\end{pgfscope}%
\begin{pgfscope}%
\pgfpathrectangle{\pgfqpoint{0.510806in}{0.456793in}}{\pgfqpoint{4.447524in}{2.201537in}}%
\pgfusepath{clip}%
\pgfsetbuttcap%
\pgfsetroundjoin%
\pgfsetlinewidth{0.803000pt}%
\definecolor{currentstroke}{rgb}{0.690196,0.690196,0.690196}%
\pgfsetstrokecolor{currentstroke}%
\pgfsetdash{{0.800000pt}{1.320000pt}}{0.000000pt}%
\pgfpathmoveto{\pgfqpoint{2.734568in}{0.456793in}}%
\pgfpathlineto{\pgfqpoint{2.734568in}{2.658330in}}%
\pgfusepath{stroke}%
\end{pgfscope}%
\begin{pgfscope}%
\pgfsetbuttcap%
\pgfsetroundjoin%
\definecolor{currentfill}{rgb}{0.000000,0.000000,0.000000}%
\pgfsetfillcolor{currentfill}%
\pgfsetlinewidth{0.803000pt}%
\definecolor{currentstroke}{rgb}{0.000000,0.000000,0.000000}%
\pgfsetstrokecolor{currentstroke}%
\pgfsetdash{}{0pt}%
\pgfsys@defobject{currentmarker}{\pgfqpoint{0.000000in}{-0.048611in}}{\pgfqpoint{0.000000in}{0.000000in}}{%
\pgfpathmoveto{\pgfqpoint{0.000000in}{0.000000in}}%
\pgfpathlineto{\pgfqpoint{0.000000in}{-0.048611in}}%
\pgfusepath{stroke,fill}%
}%
\begin{pgfscope}%
\pgfsys@transformshift{2.734568in}{0.456793in}%
\pgfsys@useobject{currentmarker}{}%
\end{pgfscope}%
\end{pgfscope}%
\begin{pgfscope}%
\definecolor{textcolor}{rgb}{0.000000,0.000000,0.000000}%
\pgfsetstrokecolor{textcolor}%
\pgfsetfillcolor{textcolor}%
\pgftext[x=2.734568in,y=0.359571in,,top]{\color{textcolor}\rmfamily\fontsize{10.000000}{12.000000}\selectfont 256}%
\end{pgfscope}%
\begin{pgfscope}%
\pgfpathrectangle{\pgfqpoint{0.510806in}{0.456793in}}{\pgfqpoint{4.447524in}{2.201537in}}%
\pgfusepath{clip}%
\pgfsetbuttcap%
\pgfsetroundjoin%
\pgfsetlinewidth{0.803000pt}%
\definecolor{currentstroke}{rgb}{0.690196,0.690196,0.690196}%
\pgfsetstrokecolor{currentstroke}%
\pgfsetdash{{0.800000pt}{1.320000pt}}{0.000000pt}%
\pgfpathmoveto{\pgfqpoint{3.846449in}{0.456793in}}%
\pgfpathlineto{\pgfqpoint{3.846449in}{2.658330in}}%
\pgfusepath{stroke}%
\end{pgfscope}%
\begin{pgfscope}%
\pgfsetbuttcap%
\pgfsetroundjoin%
\definecolor{currentfill}{rgb}{0.000000,0.000000,0.000000}%
\pgfsetfillcolor{currentfill}%
\pgfsetlinewidth{0.803000pt}%
\definecolor{currentstroke}{rgb}{0.000000,0.000000,0.000000}%
\pgfsetstrokecolor{currentstroke}%
\pgfsetdash{}{0pt}%
\pgfsys@defobject{currentmarker}{\pgfqpoint{0.000000in}{-0.048611in}}{\pgfqpoint{0.000000in}{0.000000in}}{%
\pgfpathmoveto{\pgfqpoint{0.000000in}{0.000000in}}%
\pgfpathlineto{\pgfqpoint{0.000000in}{-0.048611in}}%
\pgfusepath{stroke,fill}%
}%
\begin{pgfscope}%
\pgfsys@transformshift{3.846449in}{0.456793in}%
\pgfsys@useobject{currentmarker}{}%
\end{pgfscope}%
\end{pgfscope}%
\begin{pgfscope}%
\definecolor{textcolor}{rgb}{0.000000,0.000000,0.000000}%
\pgfsetstrokecolor{textcolor}%
\pgfsetfillcolor{textcolor}%
\pgftext[x=3.846449in,y=0.359571in,,top]{\color{textcolor}\rmfamily\fontsize{10.000000}{12.000000}\selectfont 1024}%
\end{pgfscope}%
\begin{pgfscope}%
\definecolor{textcolor}{rgb}{0.000000,0.000000,0.000000}%
\pgfsetstrokecolor{textcolor}%
\pgfsetfillcolor{textcolor}%
\pgftext[x=2.734568in,y=0.180559in,,top]{\color{textcolor}\rmfamily\fontsize{10.000000}{12.000000}\selectfont Payload (bytes)}%
\end{pgfscope}%
\begin{pgfscope}%
\pgfpathrectangle{\pgfqpoint{0.510806in}{0.456793in}}{\pgfqpoint{4.447524in}{2.201537in}}%
\pgfusepath{clip}%
\pgfsetbuttcap%
\pgfsetroundjoin%
\pgfsetlinewidth{0.803000pt}%
\definecolor{currentstroke}{rgb}{0.690196,0.690196,0.690196}%
\pgfsetstrokecolor{currentstroke}%
\pgfsetdash{{0.800000pt}{1.320000pt}}{0.000000pt}%
\pgfpathmoveto{\pgfqpoint{0.510806in}{0.676947in}}%
\pgfpathlineto{\pgfqpoint{4.958330in}{0.676947in}}%
\pgfusepath{stroke}%
\end{pgfscope}%
\begin{pgfscope}%
\pgfsetbuttcap%
\pgfsetroundjoin%
\definecolor{currentfill}{rgb}{0.000000,0.000000,0.000000}%
\pgfsetfillcolor{currentfill}%
\pgfsetlinewidth{0.803000pt}%
\definecolor{currentstroke}{rgb}{0.000000,0.000000,0.000000}%
\pgfsetstrokecolor{currentstroke}%
\pgfsetdash{}{0pt}%
\pgfsys@defobject{currentmarker}{\pgfqpoint{-0.048611in}{0.000000in}}{\pgfqpoint{-0.000000in}{0.000000in}}{%
\pgfpathmoveto{\pgfqpoint{-0.000000in}{0.000000in}}%
\pgfpathlineto{\pgfqpoint{-0.048611in}{0.000000in}}%
\pgfusepath{stroke,fill}%
}%
\begin{pgfscope}%
\pgfsys@transformshift{0.510806in}{0.676947in}%
\pgfsys@useobject{currentmarker}{}%
\end{pgfscope}%
\end{pgfscope}%
\begin{pgfscope}%
\definecolor{textcolor}{rgb}{0.000000,0.000000,0.000000}%
\pgfsetstrokecolor{textcolor}%
\pgfsetfillcolor{textcolor}%
\pgftext[x=0.236114in, y=0.628722in, left, base]{\color{textcolor}\rmfamily\fontsize{10.000000}{12.000000}\selectfont \(\displaystyle {\ensuremath{-}4}\)}%
\end{pgfscope}%
\begin{pgfscope}%
\pgfpathrectangle{\pgfqpoint{0.510806in}{0.456793in}}{\pgfqpoint{4.447524in}{2.201537in}}%
\pgfusepath{clip}%
\pgfsetbuttcap%
\pgfsetroundjoin%
\pgfsetlinewidth{0.803000pt}%
\definecolor{currentstroke}{rgb}{0.690196,0.690196,0.690196}%
\pgfsetstrokecolor{currentstroke}%
\pgfsetdash{{0.800000pt}{1.320000pt}}{0.000000pt}%
\pgfpathmoveto{\pgfqpoint{0.510806in}{1.117254in}}%
\pgfpathlineto{\pgfqpoint{4.958330in}{1.117254in}}%
\pgfusepath{stroke}%
\end{pgfscope}%
\begin{pgfscope}%
\pgfsetbuttcap%
\pgfsetroundjoin%
\definecolor{currentfill}{rgb}{0.000000,0.000000,0.000000}%
\pgfsetfillcolor{currentfill}%
\pgfsetlinewidth{0.803000pt}%
\definecolor{currentstroke}{rgb}{0.000000,0.000000,0.000000}%
\pgfsetstrokecolor{currentstroke}%
\pgfsetdash{}{0pt}%
\pgfsys@defobject{currentmarker}{\pgfqpoint{-0.048611in}{0.000000in}}{\pgfqpoint{-0.000000in}{0.000000in}}{%
\pgfpathmoveto{\pgfqpoint{-0.000000in}{0.000000in}}%
\pgfpathlineto{\pgfqpoint{-0.048611in}{0.000000in}}%
\pgfusepath{stroke,fill}%
}%
\begin{pgfscope}%
\pgfsys@transformshift{0.510806in}{1.117254in}%
\pgfsys@useobject{currentmarker}{}%
\end{pgfscope}%
\end{pgfscope}%
\begin{pgfscope}%
\definecolor{textcolor}{rgb}{0.000000,0.000000,0.000000}%
\pgfsetstrokecolor{textcolor}%
\pgfsetfillcolor{textcolor}%
\pgftext[x=0.236114in, y=1.069029in, left, base]{\color{textcolor}\rmfamily\fontsize{10.000000}{12.000000}\selectfont \(\displaystyle {\ensuremath{-}2}\)}%
\end{pgfscope}%
\begin{pgfscope}%
\pgfpathrectangle{\pgfqpoint{0.510806in}{0.456793in}}{\pgfqpoint{4.447524in}{2.201537in}}%
\pgfusepath{clip}%
\pgfsetbuttcap%
\pgfsetroundjoin%
\pgfsetlinewidth{0.803000pt}%
\definecolor{currentstroke}{rgb}{0.690196,0.690196,0.690196}%
\pgfsetstrokecolor{currentstroke}%
\pgfsetdash{{0.800000pt}{1.320000pt}}{0.000000pt}%
\pgfpathmoveto{\pgfqpoint{0.510806in}{1.557562in}}%
\pgfpathlineto{\pgfqpoint{4.958330in}{1.557562in}}%
\pgfusepath{stroke}%
\end{pgfscope}%
\begin{pgfscope}%
\pgfsetbuttcap%
\pgfsetroundjoin%
\definecolor{currentfill}{rgb}{0.000000,0.000000,0.000000}%
\pgfsetfillcolor{currentfill}%
\pgfsetlinewidth{0.803000pt}%
\definecolor{currentstroke}{rgb}{0.000000,0.000000,0.000000}%
\pgfsetstrokecolor{currentstroke}%
\pgfsetdash{}{0pt}%
\pgfsys@defobject{currentmarker}{\pgfqpoint{-0.048611in}{0.000000in}}{\pgfqpoint{-0.000000in}{0.000000in}}{%
\pgfpathmoveto{\pgfqpoint{-0.000000in}{0.000000in}}%
\pgfpathlineto{\pgfqpoint{-0.048611in}{0.000000in}}%
\pgfusepath{stroke,fill}%
}%
\begin{pgfscope}%
\pgfsys@transformshift{0.510806in}{1.557562in}%
\pgfsys@useobject{currentmarker}{}%
\end{pgfscope}%
\end{pgfscope}%
\begin{pgfscope}%
\definecolor{textcolor}{rgb}{0.000000,0.000000,0.000000}%
\pgfsetstrokecolor{textcolor}%
\pgfsetfillcolor{textcolor}%
\pgftext[x=0.344139in, y=1.509336in, left, base]{\color{textcolor}\rmfamily\fontsize{10.000000}{12.000000}\selectfont \(\displaystyle {0}\)}%
\end{pgfscope}%
\begin{pgfscope}%
\pgfpathrectangle{\pgfqpoint{0.510806in}{0.456793in}}{\pgfqpoint{4.447524in}{2.201537in}}%
\pgfusepath{clip}%
\pgfsetbuttcap%
\pgfsetroundjoin%
\pgfsetlinewidth{0.803000pt}%
\definecolor{currentstroke}{rgb}{0.690196,0.690196,0.690196}%
\pgfsetstrokecolor{currentstroke}%
\pgfsetdash{{0.800000pt}{1.320000pt}}{0.000000pt}%
\pgfpathmoveto{\pgfqpoint{0.510806in}{1.997869in}}%
\pgfpathlineto{\pgfqpoint{4.958330in}{1.997869in}}%
\pgfusepath{stroke}%
\end{pgfscope}%
\begin{pgfscope}%
\pgfsetbuttcap%
\pgfsetroundjoin%
\definecolor{currentfill}{rgb}{0.000000,0.000000,0.000000}%
\pgfsetfillcolor{currentfill}%
\pgfsetlinewidth{0.803000pt}%
\definecolor{currentstroke}{rgb}{0.000000,0.000000,0.000000}%
\pgfsetstrokecolor{currentstroke}%
\pgfsetdash{}{0pt}%
\pgfsys@defobject{currentmarker}{\pgfqpoint{-0.048611in}{0.000000in}}{\pgfqpoint{-0.000000in}{0.000000in}}{%
\pgfpathmoveto{\pgfqpoint{-0.000000in}{0.000000in}}%
\pgfpathlineto{\pgfqpoint{-0.048611in}{0.000000in}}%
\pgfusepath{stroke,fill}%
}%
\begin{pgfscope}%
\pgfsys@transformshift{0.510806in}{1.997869in}%
\pgfsys@useobject{currentmarker}{}%
\end{pgfscope}%
\end{pgfscope}%
\begin{pgfscope}%
\definecolor{textcolor}{rgb}{0.000000,0.000000,0.000000}%
\pgfsetstrokecolor{textcolor}%
\pgfsetfillcolor{textcolor}%
\pgftext[x=0.344139in, y=1.949644in, left, base]{\color{textcolor}\rmfamily\fontsize{10.000000}{12.000000}\selectfont \(\displaystyle {2}\)}%
\end{pgfscope}%
\begin{pgfscope}%
\pgfpathrectangle{\pgfqpoint{0.510806in}{0.456793in}}{\pgfqpoint{4.447524in}{2.201537in}}%
\pgfusepath{clip}%
\pgfsetbuttcap%
\pgfsetroundjoin%
\pgfsetlinewidth{0.803000pt}%
\definecolor{currentstroke}{rgb}{0.690196,0.690196,0.690196}%
\pgfsetstrokecolor{currentstroke}%
\pgfsetdash{{0.800000pt}{1.320000pt}}{0.000000pt}%
\pgfpathmoveto{\pgfqpoint{0.510806in}{2.438176in}}%
\pgfpathlineto{\pgfqpoint{4.958330in}{2.438176in}}%
\pgfusepath{stroke}%
\end{pgfscope}%
\begin{pgfscope}%
\pgfsetbuttcap%
\pgfsetroundjoin%
\definecolor{currentfill}{rgb}{0.000000,0.000000,0.000000}%
\pgfsetfillcolor{currentfill}%
\pgfsetlinewidth{0.803000pt}%
\definecolor{currentstroke}{rgb}{0.000000,0.000000,0.000000}%
\pgfsetstrokecolor{currentstroke}%
\pgfsetdash{}{0pt}%
\pgfsys@defobject{currentmarker}{\pgfqpoint{-0.048611in}{0.000000in}}{\pgfqpoint{-0.000000in}{0.000000in}}{%
\pgfpathmoveto{\pgfqpoint{-0.000000in}{0.000000in}}%
\pgfpathlineto{\pgfqpoint{-0.048611in}{0.000000in}}%
\pgfusepath{stroke,fill}%
}%
\begin{pgfscope}%
\pgfsys@transformshift{0.510806in}{2.438176in}%
\pgfsys@useobject{currentmarker}{}%
\end{pgfscope}%
\end{pgfscope}%
\begin{pgfscope}%
\definecolor{textcolor}{rgb}{0.000000,0.000000,0.000000}%
\pgfsetstrokecolor{textcolor}%
\pgfsetfillcolor{textcolor}%
\pgftext[x=0.344139in, y=2.389951in, left, base]{\color{textcolor}\rmfamily\fontsize{10.000000}{12.000000}\selectfont \(\displaystyle {4}\)}%
\end{pgfscope}%
\begin{pgfscope}%
\definecolor{textcolor}{rgb}{0.000000,0.000000,0.000000}%
\pgfsetstrokecolor{textcolor}%
\pgfsetfillcolor{textcolor}%
\pgftext[x=0.180559in,y=1.557562in,,bottom,rotate=90.000000]{\color{textcolor}\rmfamily\fontsize{10.000000}{12.000000}\selectfont Jitter (\(\displaystyle \mu\)s)}%
\end{pgfscope}%
\begin{pgfscope}%
\pgfpathrectangle{\pgfqpoint{0.510806in}{0.456793in}}{\pgfqpoint{4.447524in}{2.201537in}}%
\pgfusepath{clip}%
\pgfsetbuttcap%
\pgfsetmiterjoin%
\definecolor{currentfill}{rgb}{0.400000,0.760784,0.647059}%
\pgfsetfillcolor{currentfill}%
\pgfsetlinewidth{1.003750pt}%
\definecolor{currentstroke}{rgb}{0.000000,0.000000,0.000000}%
\pgfsetstrokecolor{currentstroke}%
\pgfsetdash{}{0pt}%
\pgfpathmoveto{\pgfqpoint{1.177935in}{1.438238in}}%
\pgfpathlineto{\pgfqpoint{1.444786in}{1.438238in}}%
\pgfpathlineto{\pgfqpoint{1.444786in}{1.553141in}}%
\pgfpathlineto{\pgfqpoint{1.378073in}{1.554480in}}%
\pgfpathlineto{\pgfqpoint{1.444786in}{1.555818in}}%
\pgfpathlineto{\pgfqpoint{1.444786in}{1.707927in}}%
\pgfpathlineto{\pgfqpoint{1.177935in}{1.707927in}}%
\pgfpathlineto{\pgfqpoint{1.177935in}{1.555818in}}%
\pgfpathlineto{\pgfqpoint{1.244648in}{1.554480in}}%
\pgfpathlineto{\pgfqpoint{1.177935in}{1.553141in}}%
\pgfpathlineto{\pgfqpoint{1.177935in}{1.438238in}}%
\pgfpathlineto{\pgfqpoint{1.177935in}{1.438238in}}%
\pgfpathclose%
\pgfusepath{stroke,fill}%
\end{pgfscope}%
\begin{pgfscope}%
\pgfpathrectangle{\pgfqpoint{0.510806in}{0.456793in}}{\pgfqpoint{4.447524in}{2.201537in}}%
\pgfusepath{clip}%
\pgfsetrectcap%
\pgfsetroundjoin%
\pgfsetlinewidth{1.003750pt}%
\definecolor{currentstroke}{rgb}{0.000000,0.000000,0.000000}%
\pgfsetstrokecolor{currentstroke}%
\pgfsetdash{}{0pt}%
\pgfpathmoveto{\pgfqpoint{1.311361in}{1.438238in}}%
\pgfpathlineto{\pgfqpoint{1.311361in}{1.033816in}}%
\pgfusepath{stroke}%
\end{pgfscope}%
\begin{pgfscope}%
\pgfpathrectangle{\pgfqpoint{0.510806in}{0.456793in}}{\pgfqpoint{4.447524in}{2.201537in}}%
\pgfusepath{clip}%
\pgfsetrectcap%
\pgfsetroundjoin%
\pgfsetlinewidth{1.003750pt}%
\definecolor{currentstroke}{rgb}{0.000000,0.000000,0.000000}%
\pgfsetstrokecolor{currentstroke}%
\pgfsetdash{}{0pt}%
\pgfpathmoveto{\pgfqpoint{1.311361in}{1.707927in}}%
\pgfpathlineto{\pgfqpoint{1.311361in}{2.112129in}}%
\pgfusepath{stroke}%
\end{pgfscope}%
\begin{pgfscope}%
\pgfpathrectangle{\pgfqpoint{0.510806in}{0.456793in}}{\pgfqpoint{4.447524in}{2.201537in}}%
\pgfusepath{clip}%
\pgfsetrectcap%
\pgfsetroundjoin%
\pgfsetlinewidth{1.003750pt}%
\definecolor{currentstroke}{rgb}{0.000000,0.000000,0.000000}%
\pgfsetstrokecolor{currentstroke}%
\pgfsetdash{}{0pt}%
\pgfpathmoveto{\pgfqpoint{1.244648in}{1.033816in}}%
\pgfpathlineto{\pgfqpoint{1.378073in}{1.033816in}}%
\pgfusepath{stroke}%
\end{pgfscope}%
\begin{pgfscope}%
\pgfpathrectangle{\pgfqpoint{0.510806in}{0.456793in}}{\pgfqpoint{4.447524in}{2.201537in}}%
\pgfusepath{clip}%
\pgfsetrectcap%
\pgfsetroundjoin%
\pgfsetlinewidth{1.003750pt}%
\definecolor{currentstroke}{rgb}{0.000000,0.000000,0.000000}%
\pgfsetstrokecolor{currentstroke}%
\pgfsetdash{}{0pt}%
\pgfpathmoveto{\pgfqpoint{1.244648in}{2.112129in}}%
\pgfpathlineto{\pgfqpoint{1.378073in}{2.112129in}}%
\pgfusepath{stroke}%
\end{pgfscope}%
\begin{pgfscope}%
\pgfpathrectangle{\pgfqpoint{0.510806in}{0.456793in}}{\pgfqpoint{4.447524in}{2.201537in}}%
\pgfusepath{clip}%
\pgfsetbuttcap%
\pgfsetmiterjoin%
\definecolor{currentfill}{rgb}{0.400000,0.760784,0.647059}%
\pgfsetfillcolor{currentfill}%
\pgfsetlinewidth{1.003750pt}%
\definecolor{currentstroke}{rgb}{0.000000,0.000000,0.000000}%
\pgfsetstrokecolor{currentstroke}%
\pgfsetdash{}{0pt}%
\pgfpathmoveto{\pgfqpoint{2.289816in}{1.444843in}}%
\pgfpathlineto{\pgfqpoint{2.556667in}{1.444843in}}%
\pgfpathlineto{\pgfqpoint{2.556667in}{1.549555in}}%
\pgfpathlineto{\pgfqpoint{2.489954in}{1.550737in}}%
\pgfpathlineto{\pgfqpoint{2.556667in}{1.551918in}}%
\pgfpathlineto{\pgfqpoint{2.556667in}{1.682829in}}%
\pgfpathlineto{\pgfqpoint{2.289816in}{1.682829in}}%
\pgfpathlineto{\pgfqpoint{2.289816in}{1.551918in}}%
\pgfpathlineto{\pgfqpoint{2.356529in}{1.550737in}}%
\pgfpathlineto{\pgfqpoint{2.289816in}{1.549555in}}%
\pgfpathlineto{\pgfqpoint{2.289816in}{1.444843in}}%
\pgfpathlineto{\pgfqpoint{2.289816in}{1.444843in}}%
\pgfpathclose%
\pgfusepath{stroke,fill}%
\end{pgfscope}%
\begin{pgfscope}%
\pgfpathrectangle{\pgfqpoint{0.510806in}{0.456793in}}{\pgfqpoint{4.447524in}{2.201537in}}%
\pgfusepath{clip}%
\pgfsetrectcap%
\pgfsetroundjoin%
\pgfsetlinewidth{1.003750pt}%
\definecolor{currentstroke}{rgb}{0.000000,0.000000,0.000000}%
\pgfsetstrokecolor{currentstroke}%
\pgfsetdash{}{0pt}%
\pgfpathmoveto{\pgfqpoint{2.423242in}{1.444843in}}%
\pgfpathlineto{\pgfqpoint{2.423242in}{1.087974in}}%
\pgfusepath{stroke}%
\end{pgfscope}%
\begin{pgfscope}%
\pgfpathrectangle{\pgfqpoint{0.510806in}{0.456793in}}{\pgfqpoint{4.447524in}{2.201537in}}%
\pgfusepath{clip}%
\pgfsetrectcap%
\pgfsetroundjoin%
\pgfsetlinewidth{1.003750pt}%
\definecolor{currentstroke}{rgb}{0.000000,0.000000,0.000000}%
\pgfsetstrokecolor{currentstroke}%
\pgfsetdash{}{0pt}%
\pgfpathmoveto{\pgfqpoint{2.423242in}{1.682829in}}%
\pgfpathlineto{\pgfqpoint{2.423242in}{2.039258in}}%
\pgfusepath{stroke}%
\end{pgfscope}%
\begin{pgfscope}%
\pgfpathrectangle{\pgfqpoint{0.510806in}{0.456793in}}{\pgfqpoint{4.447524in}{2.201537in}}%
\pgfusepath{clip}%
\pgfsetrectcap%
\pgfsetroundjoin%
\pgfsetlinewidth{1.003750pt}%
\definecolor{currentstroke}{rgb}{0.000000,0.000000,0.000000}%
\pgfsetstrokecolor{currentstroke}%
\pgfsetdash{}{0pt}%
\pgfpathmoveto{\pgfqpoint{2.356529in}{1.087974in}}%
\pgfpathlineto{\pgfqpoint{2.489954in}{1.087974in}}%
\pgfusepath{stroke}%
\end{pgfscope}%
\begin{pgfscope}%
\pgfpathrectangle{\pgfqpoint{0.510806in}{0.456793in}}{\pgfqpoint{4.447524in}{2.201537in}}%
\pgfusepath{clip}%
\pgfsetrectcap%
\pgfsetroundjoin%
\pgfsetlinewidth{1.003750pt}%
\definecolor{currentstroke}{rgb}{0.000000,0.000000,0.000000}%
\pgfsetstrokecolor{currentstroke}%
\pgfsetdash{}{0pt}%
\pgfpathmoveto{\pgfqpoint{2.356529in}{2.039258in}}%
\pgfpathlineto{\pgfqpoint{2.489954in}{2.039258in}}%
\pgfusepath{stroke}%
\end{pgfscope}%
\begin{pgfscope}%
\pgfpathrectangle{\pgfqpoint{0.510806in}{0.456793in}}{\pgfqpoint{4.447524in}{2.201537in}}%
\pgfusepath{clip}%
\pgfsetbuttcap%
\pgfsetmiterjoin%
\definecolor{currentfill}{rgb}{0.400000,0.760784,0.647059}%
\pgfsetfillcolor{currentfill}%
\pgfsetlinewidth{1.003750pt}%
\definecolor{currentstroke}{rgb}{0.000000,0.000000,0.000000}%
\pgfsetstrokecolor{currentstroke}%
\pgfsetdash{}{0pt}%
\pgfpathmoveto{\pgfqpoint{3.401697in}{1.452548in}}%
\pgfpathlineto{\pgfqpoint{3.668548in}{1.452548in}}%
\pgfpathlineto{\pgfqpoint{3.668548in}{1.536788in}}%
\pgfpathlineto{\pgfqpoint{3.601835in}{1.537968in}}%
\pgfpathlineto{\pgfqpoint{3.668548in}{1.539148in}}%
\pgfpathlineto{\pgfqpoint{3.668548in}{1.690314in}}%
\pgfpathlineto{\pgfqpoint{3.401697in}{1.690314in}}%
\pgfpathlineto{\pgfqpoint{3.401697in}{1.539148in}}%
\pgfpathlineto{\pgfqpoint{3.468410in}{1.537968in}}%
\pgfpathlineto{\pgfqpoint{3.401697in}{1.536788in}}%
\pgfpathlineto{\pgfqpoint{3.401697in}{1.452548in}}%
\pgfpathlineto{\pgfqpoint{3.401697in}{1.452548in}}%
\pgfpathclose%
\pgfusepath{stroke,fill}%
\end{pgfscope}%
\begin{pgfscope}%
\pgfpathrectangle{\pgfqpoint{0.510806in}{0.456793in}}{\pgfqpoint{4.447524in}{2.201537in}}%
\pgfusepath{clip}%
\pgfsetrectcap%
\pgfsetroundjoin%
\pgfsetlinewidth{1.003750pt}%
\definecolor{currentstroke}{rgb}{0.000000,0.000000,0.000000}%
\pgfsetstrokecolor{currentstroke}%
\pgfsetdash{}{0pt}%
\pgfpathmoveto{\pgfqpoint{3.535122in}{1.452548in}}%
\pgfpathlineto{\pgfqpoint{3.535122in}{1.095899in}}%
\pgfusepath{stroke}%
\end{pgfscope}%
\begin{pgfscope}%
\pgfpathrectangle{\pgfqpoint{0.510806in}{0.456793in}}{\pgfqpoint{4.447524in}{2.201537in}}%
\pgfusepath{clip}%
\pgfsetrectcap%
\pgfsetroundjoin%
\pgfsetlinewidth{1.003750pt}%
\definecolor{currentstroke}{rgb}{0.000000,0.000000,0.000000}%
\pgfsetstrokecolor{currentstroke}%
\pgfsetdash{}{0pt}%
\pgfpathmoveto{\pgfqpoint{3.535122in}{1.690314in}}%
\pgfpathlineto{\pgfqpoint{3.535122in}{2.046743in}}%
\pgfusepath{stroke}%
\end{pgfscope}%
\begin{pgfscope}%
\pgfpathrectangle{\pgfqpoint{0.510806in}{0.456793in}}{\pgfqpoint{4.447524in}{2.201537in}}%
\pgfusepath{clip}%
\pgfsetrectcap%
\pgfsetroundjoin%
\pgfsetlinewidth{1.003750pt}%
\definecolor{currentstroke}{rgb}{0.000000,0.000000,0.000000}%
\pgfsetstrokecolor{currentstroke}%
\pgfsetdash{}{0pt}%
\pgfpathmoveto{\pgfqpoint{3.468410in}{1.095899in}}%
\pgfpathlineto{\pgfqpoint{3.601835in}{1.095899in}}%
\pgfusepath{stroke}%
\end{pgfscope}%
\begin{pgfscope}%
\pgfpathrectangle{\pgfqpoint{0.510806in}{0.456793in}}{\pgfqpoint{4.447524in}{2.201537in}}%
\pgfusepath{clip}%
\pgfsetrectcap%
\pgfsetroundjoin%
\pgfsetlinewidth{1.003750pt}%
\definecolor{currentstroke}{rgb}{0.000000,0.000000,0.000000}%
\pgfsetstrokecolor{currentstroke}%
\pgfsetdash{}{0pt}%
\pgfpathmoveto{\pgfqpoint{3.468410in}{2.046743in}}%
\pgfpathlineto{\pgfqpoint{3.601835in}{2.046743in}}%
\pgfusepath{stroke}%
\end{pgfscope}%
\begin{pgfscope}%
\pgfpathrectangle{\pgfqpoint{0.510806in}{0.456793in}}{\pgfqpoint{4.447524in}{2.201537in}}%
\pgfusepath{clip}%
\pgfsetbuttcap%
\pgfsetmiterjoin%
\definecolor{currentfill}{rgb}{0.988235,0.552941,0.384314}%
\pgfsetfillcolor{currentfill}%
\pgfsetlinewidth{1.003750pt}%
\definecolor{currentstroke}{rgb}{0.000000,0.000000,0.000000}%
\pgfsetstrokecolor{currentstroke}%
\pgfsetdash{}{0pt}%
\pgfpathmoveto{\pgfqpoint{1.489262in}{1.447045in}}%
\pgfpathlineto{\pgfqpoint{1.756113in}{1.447045in}}%
\pgfpathlineto{\pgfqpoint{1.756113in}{1.545436in}}%
\pgfpathlineto{\pgfqpoint{1.689400in}{1.546554in}}%
\pgfpathlineto{\pgfqpoint{1.756113in}{1.547672in}}%
\pgfpathlineto{\pgfqpoint{1.756113in}{1.672262in}}%
\pgfpathlineto{\pgfqpoint{1.489262in}{1.672262in}}%
\pgfpathlineto{\pgfqpoint{1.489262in}{1.547672in}}%
\pgfpathlineto{\pgfqpoint{1.555974in}{1.546554in}}%
\pgfpathlineto{\pgfqpoint{1.489262in}{1.545436in}}%
\pgfpathlineto{\pgfqpoint{1.489262in}{1.447045in}}%
\pgfpathlineto{\pgfqpoint{1.489262in}{1.447045in}}%
\pgfpathclose%
\pgfusepath{stroke,fill}%
\end{pgfscope}%
\begin{pgfscope}%
\pgfpathrectangle{\pgfqpoint{0.510806in}{0.456793in}}{\pgfqpoint{4.447524in}{2.201537in}}%
\pgfusepath{clip}%
\pgfsetrectcap%
\pgfsetroundjoin%
\pgfsetlinewidth{1.003750pt}%
\definecolor{currentstroke}{rgb}{0.000000,0.000000,0.000000}%
\pgfsetstrokecolor{currentstroke}%
\pgfsetdash{}{0pt}%
\pgfpathmoveto{\pgfqpoint{1.622687in}{1.447045in}}%
\pgfpathlineto{\pgfqpoint{1.622687in}{1.109329in}}%
\pgfusepath{stroke}%
\end{pgfscope}%
\begin{pgfscope}%
\pgfpathrectangle{\pgfqpoint{0.510806in}{0.456793in}}{\pgfqpoint{4.447524in}{2.201537in}}%
\pgfusepath{clip}%
\pgfsetrectcap%
\pgfsetroundjoin%
\pgfsetlinewidth{1.003750pt}%
\definecolor{currentstroke}{rgb}{0.000000,0.000000,0.000000}%
\pgfsetstrokecolor{currentstroke}%
\pgfsetdash{}{0pt}%
\pgfpathmoveto{\pgfqpoint{1.622687in}{1.672262in}}%
\pgfpathlineto{\pgfqpoint{1.622687in}{2.009977in}}%
\pgfusepath{stroke}%
\end{pgfscope}%
\begin{pgfscope}%
\pgfpathrectangle{\pgfqpoint{0.510806in}{0.456793in}}{\pgfqpoint{4.447524in}{2.201537in}}%
\pgfusepath{clip}%
\pgfsetrectcap%
\pgfsetroundjoin%
\pgfsetlinewidth{1.003750pt}%
\definecolor{currentstroke}{rgb}{0.000000,0.000000,0.000000}%
\pgfsetstrokecolor{currentstroke}%
\pgfsetdash{}{0pt}%
\pgfpathmoveto{\pgfqpoint{1.555974in}{1.109329in}}%
\pgfpathlineto{\pgfqpoint{1.689400in}{1.109329in}}%
\pgfusepath{stroke}%
\end{pgfscope}%
\begin{pgfscope}%
\pgfpathrectangle{\pgfqpoint{0.510806in}{0.456793in}}{\pgfqpoint{4.447524in}{2.201537in}}%
\pgfusepath{clip}%
\pgfsetrectcap%
\pgfsetroundjoin%
\pgfsetlinewidth{1.003750pt}%
\definecolor{currentstroke}{rgb}{0.000000,0.000000,0.000000}%
\pgfsetstrokecolor{currentstroke}%
\pgfsetdash{}{0pt}%
\pgfpathmoveto{\pgfqpoint{1.555974in}{2.009977in}}%
\pgfpathlineto{\pgfqpoint{1.689400in}{2.009977in}}%
\pgfusepath{stroke}%
\end{pgfscope}%
\begin{pgfscope}%
\pgfpathrectangle{\pgfqpoint{0.510806in}{0.456793in}}{\pgfqpoint{4.447524in}{2.201537in}}%
\pgfusepath{clip}%
\pgfsetbuttcap%
\pgfsetmiterjoin%
\definecolor{currentfill}{rgb}{0.988235,0.552941,0.384314}%
\pgfsetfillcolor{currentfill}%
\pgfsetlinewidth{1.003750pt}%
\definecolor{currentstroke}{rgb}{0.000000,0.000000,0.000000}%
\pgfsetstrokecolor{currentstroke}%
\pgfsetdash{}{0pt}%
\pgfpathmoveto{\pgfqpoint{2.601142in}{1.443852in}}%
\pgfpathlineto{\pgfqpoint{2.867994in}{1.443852in}}%
\pgfpathlineto{\pgfqpoint{2.867994in}{1.553353in}}%
\pgfpathlineto{\pgfqpoint{2.801281in}{1.554480in}}%
\pgfpathlineto{\pgfqpoint{2.867994in}{1.555606in}}%
\pgfpathlineto{\pgfqpoint{2.867994in}{1.670721in}}%
\pgfpathlineto{\pgfqpoint{2.601142in}{1.670721in}}%
\pgfpathlineto{\pgfqpoint{2.601142in}{1.555606in}}%
\pgfpathlineto{\pgfqpoint{2.667855in}{1.554480in}}%
\pgfpathlineto{\pgfqpoint{2.601142in}{1.553353in}}%
\pgfpathlineto{\pgfqpoint{2.601142in}{1.443852in}}%
\pgfpathlineto{\pgfqpoint{2.601142in}{1.443852in}}%
\pgfpathclose%
\pgfusepath{stroke,fill}%
\end{pgfscope}%
\begin{pgfscope}%
\pgfpathrectangle{\pgfqpoint{0.510806in}{0.456793in}}{\pgfqpoint{4.447524in}{2.201537in}}%
\pgfusepath{clip}%
\pgfsetrectcap%
\pgfsetroundjoin%
\pgfsetlinewidth{1.003750pt}%
\definecolor{currentstroke}{rgb}{0.000000,0.000000,0.000000}%
\pgfsetstrokecolor{currentstroke}%
\pgfsetdash{}{0pt}%
\pgfpathmoveto{\pgfqpoint{2.734568in}{1.443852in}}%
\pgfpathlineto{\pgfqpoint{2.734568in}{1.103825in}}%
\pgfusepath{stroke}%
\end{pgfscope}%
\begin{pgfscope}%
\pgfpathrectangle{\pgfqpoint{0.510806in}{0.456793in}}{\pgfqpoint{4.447524in}{2.201537in}}%
\pgfusepath{clip}%
\pgfsetrectcap%
\pgfsetroundjoin%
\pgfsetlinewidth{1.003750pt}%
\definecolor{currentstroke}{rgb}{0.000000,0.000000,0.000000}%
\pgfsetstrokecolor{currentstroke}%
\pgfsetdash{}{0pt}%
\pgfpathmoveto{\pgfqpoint{2.734568in}{1.670721in}}%
\pgfpathlineto{\pgfqpoint{2.734568in}{2.010858in}}%
\pgfusepath{stroke}%
\end{pgfscope}%
\begin{pgfscope}%
\pgfpathrectangle{\pgfqpoint{0.510806in}{0.456793in}}{\pgfqpoint{4.447524in}{2.201537in}}%
\pgfusepath{clip}%
\pgfsetrectcap%
\pgfsetroundjoin%
\pgfsetlinewidth{1.003750pt}%
\definecolor{currentstroke}{rgb}{0.000000,0.000000,0.000000}%
\pgfsetstrokecolor{currentstroke}%
\pgfsetdash{}{0pt}%
\pgfpathmoveto{\pgfqpoint{2.667855in}{1.103825in}}%
\pgfpathlineto{\pgfqpoint{2.801281in}{1.103825in}}%
\pgfusepath{stroke}%
\end{pgfscope}%
\begin{pgfscope}%
\pgfpathrectangle{\pgfqpoint{0.510806in}{0.456793in}}{\pgfqpoint{4.447524in}{2.201537in}}%
\pgfusepath{clip}%
\pgfsetrectcap%
\pgfsetroundjoin%
\pgfsetlinewidth{1.003750pt}%
\definecolor{currentstroke}{rgb}{0.000000,0.000000,0.000000}%
\pgfsetstrokecolor{currentstroke}%
\pgfsetdash{}{0pt}%
\pgfpathmoveto{\pgfqpoint{2.667855in}{2.010858in}}%
\pgfpathlineto{\pgfqpoint{2.801281in}{2.010858in}}%
\pgfusepath{stroke}%
\end{pgfscope}%
\begin{pgfscope}%
\pgfpathrectangle{\pgfqpoint{0.510806in}{0.456793in}}{\pgfqpoint{4.447524in}{2.201537in}}%
\pgfusepath{clip}%
\pgfsetbuttcap%
\pgfsetmiterjoin%
\definecolor{currentfill}{rgb}{0.988235,0.552941,0.384314}%
\pgfsetfillcolor{currentfill}%
\pgfsetlinewidth{1.003750pt}%
\definecolor{currentstroke}{rgb}{0.000000,0.000000,0.000000}%
\pgfsetstrokecolor{currentstroke}%
\pgfsetdash{}{0pt}%
\pgfpathmoveto{\pgfqpoint{3.713023in}{1.448586in}}%
\pgfpathlineto{\pgfqpoint{3.979875in}{1.448586in}}%
\pgfpathlineto{\pgfqpoint{3.979875in}{1.549206in}}%
\pgfpathlineto{\pgfqpoint{3.913162in}{1.550297in}}%
\pgfpathlineto{\pgfqpoint{3.979875in}{1.551387in}}%
\pgfpathlineto{\pgfqpoint{3.979875in}{1.668299in}}%
\pgfpathlineto{\pgfqpoint{3.713023in}{1.668299in}}%
\pgfpathlineto{\pgfqpoint{3.713023in}{1.551387in}}%
\pgfpathlineto{\pgfqpoint{3.779736in}{1.550297in}}%
\pgfpathlineto{\pgfqpoint{3.713023in}{1.549206in}}%
\pgfpathlineto{\pgfqpoint{3.713023in}{1.448586in}}%
\pgfpathlineto{\pgfqpoint{3.713023in}{1.448586in}}%
\pgfpathclose%
\pgfusepath{stroke,fill}%
\end{pgfscope}%
\begin{pgfscope}%
\pgfpathrectangle{\pgfqpoint{0.510806in}{0.456793in}}{\pgfqpoint{4.447524in}{2.201537in}}%
\pgfusepath{clip}%
\pgfsetrectcap%
\pgfsetroundjoin%
\pgfsetlinewidth{1.003750pt}%
\definecolor{currentstroke}{rgb}{0.000000,0.000000,0.000000}%
\pgfsetstrokecolor{currentstroke}%
\pgfsetdash{}{0pt}%
\pgfpathmoveto{\pgfqpoint{3.846449in}{1.448586in}}%
\pgfpathlineto{\pgfqpoint{3.846449in}{1.119016in}}%
\pgfusepath{stroke}%
\end{pgfscope}%
\begin{pgfscope}%
\pgfpathrectangle{\pgfqpoint{0.510806in}{0.456793in}}{\pgfqpoint{4.447524in}{2.201537in}}%
\pgfusepath{clip}%
\pgfsetrectcap%
\pgfsetroundjoin%
\pgfsetlinewidth{1.003750pt}%
\definecolor{currentstroke}{rgb}{0.000000,0.000000,0.000000}%
\pgfsetstrokecolor{currentstroke}%
\pgfsetdash{}{0pt}%
\pgfpathmoveto{\pgfqpoint{3.846449in}{1.668299in}}%
\pgfpathlineto{\pgfqpoint{3.846449in}{1.997869in}}%
\pgfusepath{stroke}%
\end{pgfscope}%
\begin{pgfscope}%
\pgfpathrectangle{\pgfqpoint{0.510806in}{0.456793in}}{\pgfqpoint{4.447524in}{2.201537in}}%
\pgfusepath{clip}%
\pgfsetrectcap%
\pgfsetroundjoin%
\pgfsetlinewidth{1.003750pt}%
\definecolor{currentstroke}{rgb}{0.000000,0.000000,0.000000}%
\pgfsetstrokecolor{currentstroke}%
\pgfsetdash{}{0pt}%
\pgfpathmoveto{\pgfqpoint{3.779736in}{1.119016in}}%
\pgfpathlineto{\pgfqpoint{3.913162in}{1.119016in}}%
\pgfusepath{stroke}%
\end{pgfscope}%
\begin{pgfscope}%
\pgfpathrectangle{\pgfqpoint{0.510806in}{0.456793in}}{\pgfqpoint{4.447524in}{2.201537in}}%
\pgfusepath{clip}%
\pgfsetrectcap%
\pgfsetroundjoin%
\pgfsetlinewidth{1.003750pt}%
\definecolor{currentstroke}{rgb}{0.000000,0.000000,0.000000}%
\pgfsetstrokecolor{currentstroke}%
\pgfsetdash{}{0pt}%
\pgfpathmoveto{\pgfqpoint{3.779736in}{1.997869in}}%
\pgfpathlineto{\pgfqpoint{3.913162in}{1.997869in}}%
\pgfusepath{stroke}%
\end{pgfscope}%
\begin{pgfscope}%
\pgfpathrectangle{\pgfqpoint{0.510806in}{0.456793in}}{\pgfqpoint{4.447524in}{2.201537in}}%
\pgfusepath{clip}%
\pgfsetbuttcap%
\pgfsetmiterjoin%
\definecolor{currentfill}{rgb}{0.552941,0.627451,0.796078}%
\pgfsetfillcolor{currentfill}%
\pgfsetlinewidth{1.003750pt}%
\definecolor{currentstroke}{rgb}{0.000000,0.000000,0.000000}%
\pgfsetstrokecolor{currentstroke}%
\pgfsetdash{}{0pt}%
\pgfpathmoveto{\pgfqpoint{1.800588in}{1.444623in}}%
\pgfpathlineto{\pgfqpoint{2.067440in}{1.444623in}}%
\pgfpathlineto{\pgfqpoint{2.067440in}{1.550487in}}%
\pgfpathlineto{\pgfqpoint{2.000727in}{1.551618in}}%
\pgfpathlineto{\pgfqpoint{2.067440in}{1.552748in}}%
\pgfpathlineto{\pgfqpoint{2.067440in}{1.672262in}}%
\pgfpathlineto{\pgfqpoint{1.800588in}{1.672262in}}%
\pgfpathlineto{\pgfqpoint{1.800588in}{1.552748in}}%
\pgfpathlineto{\pgfqpoint{1.867301in}{1.551618in}}%
\pgfpathlineto{\pgfqpoint{1.800588in}{1.550487in}}%
\pgfpathlineto{\pgfqpoint{1.800588in}{1.444623in}}%
\pgfpathlineto{\pgfqpoint{1.800588in}{1.444623in}}%
\pgfpathclose%
\pgfusepath{stroke,fill}%
\end{pgfscope}%
\begin{pgfscope}%
\pgfpathrectangle{\pgfqpoint{0.510806in}{0.456793in}}{\pgfqpoint{4.447524in}{2.201537in}}%
\pgfusepath{clip}%
\pgfsetrectcap%
\pgfsetroundjoin%
\pgfsetlinewidth{1.003750pt}%
\definecolor{currentstroke}{rgb}{0.000000,0.000000,0.000000}%
\pgfsetstrokecolor{currentstroke}%
\pgfsetdash{}{0pt}%
\pgfpathmoveto{\pgfqpoint{1.934014in}{1.444623in}}%
\pgfpathlineto{\pgfqpoint{1.934014in}{1.103164in}}%
\pgfusepath{stroke}%
\end{pgfscope}%
\begin{pgfscope}%
\pgfpathrectangle{\pgfqpoint{0.510806in}{0.456793in}}{\pgfqpoint{4.447524in}{2.201537in}}%
\pgfusepath{clip}%
\pgfsetrectcap%
\pgfsetroundjoin%
\pgfsetlinewidth{1.003750pt}%
\definecolor{currentstroke}{rgb}{0.000000,0.000000,0.000000}%
\pgfsetstrokecolor{currentstroke}%
\pgfsetdash{}{0pt}%
\pgfpathmoveto{\pgfqpoint{1.934014in}{1.672262in}}%
\pgfpathlineto{\pgfqpoint{1.934014in}{2.013720in}}%
\pgfusepath{stroke}%
\end{pgfscope}%
\begin{pgfscope}%
\pgfpathrectangle{\pgfqpoint{0.510806in}{0.456793in}}{\pgfqpoint{4.447524in}{2.201537in}}%
\pgfusepath{clip}%
\pgfsetrectcap%
\pgfsetroundjoin%
\pgfsetlinewidth{1.003750pt}%
\definecolor{currentstroke}{rgb}{0.000000,0.000000,0.000000}%
\pgfsetstrokecolor{currentstroke}%
\pgfsetdash{}{0pt}%
\pgfpathmoveto{\pgfqpoint{1.867301in}{1.103164in}}%
\pgfpathlineto{\pgfqpoint{2.000727in}{1.103164in}}%
\pgfusepath{stroke}%
\end{pgfscope}%
\begin{pgfscope}%
\pgfpathrectangle{\pgfqpoint{0.510806in}{0.456793in}}{\pgfqpoint{4.447524in}{2.201537in}}%
\pgfusepath{clip}%
\pgfsetrectcap%
\pgfsetroundjoin%
\pgfsetlinewidth{1.003750pt}%
\definecolor{currentstroke}{rgb}{0.000000,0.000000,0.000000}%
\pgfsetstrokecolor{currentstroke}%
\pgfsetdash{}{0pt}%
\pgfpathmoveto{\pgfqpoint{1.867301in}{2.013720in}}%
\pgfpathlineto{\pgfqpoint{2.000727in}{2.013720in}}%
\pgfusepath{stroke}%
\end{pgfscope}%
\begin{pgfscope}%
\pgfpathrectangle{\pgfqpoint{0.510806in}{0.456793in}}{\pgfqpoint{4.447524in}{2.201537in}}%
\pgfusepath{clip}%
\pgfsetbuttcap%
\pgfsetmiterjoin%
\definecolor{currentfill}{rgb}{0.552941,0.627451,0.796078}%
\pgfsetfillcolor{currentfill}%
\pgfsetlinewidth{1.003750pt}%
\definecolor{currentstroke}{rgb}{0.000000,0.000000,0.000000}%
\pgfsetstrokecolor{currentstroke}%
\pgfsetdash{}{0pt}%
\pgfpathmoveto{\pgfqpoint{2.912469in}{1.446164in}}%
\pgfpathlineto{\pgfqpoint{3.179321in}{1.446164in}}%
\pgfpathlineto{\pgfqpoint{3.179321in}{1.553809in}}%
\pgfpathlineto{\pgfqpoint{3.112608in}{1.554920in}}%
\pgfpathlineto{\pgfqpoint{3.179321in}{1.556030in}}%
\pgfpathlineto{\pgfqpoint{3.179321in}{1.669840in}}%
\pgfpathlineto{\pgfqpoint{2.912469in}{1.669840in}}%
\pgfpathlineto{\pgfqpoint{2.912469in}{1.556030in}}%
\pgfpathlineto{\pgfqpoint{2.979182in}{1.554920in}}%
\pgfpathlineto{\pgfqpoint{2.912469in}{1.553809in}}%
\pgfpathlineto{\pgfqpoint{2.912469in}{1.446164in}}%
\pgfpathlineto{\pgfqpoint{2.912469in}{1.446164in}}%
\pgfpathclose%
\pgfusepath{stroke,fill}%
\end{pgfscope}%
\begin{pgfscope}%
\pgfpathrectangle{\pgfqpoint{0.510806in}{0.456793in}}{\pgfqpoint{4.447524in}{2.201537in}}%
\pgfusepath{clip}%
\pgfsetrectcap%
\pgfsetroundjoin%
\pgfsetlinewidth{1.003750pt}%
\definecolor{currentstroke}{rgb}{0.000000,0.000000,0.000000}%
\pgfsetstrokecolor{currentstroke}%
\pgfsetdash{}{0pt}%
\pgfpathmoveto{\pgfqpoint{3.045895in}{1.446164in}}%
\pgfpathlineto{\pgfqpoint{3.045895in}{1.110650in}}%
\pgfusepath{stroke}%
\end{pgfscope}%
\begin{pgfscope}%
\pgfpathrectangle{\pgfqpoint{0.510806in}{0.456793in}}{\pgfqpoint{4.447524in}{2.201537in}}%
\pgfusepath{clip}%
\pgfsetrectcap%
\pgfsetroundjoin%
\pgfsetlinewidth{1.003750pt}%
\definecolor{currentstroke}{rgb}{0.000000,0.000000,0.000000}%
\pgfsetstrokecolor{currentstroke}%
\pgfsetdash{}{0pt}%
\pgfpathmoveto{\pgfqpoint{3.045895in}{1.669840in}}%
\pgfpathlineto{\pgfqpoint{3.045895in}{2.005354in}}%
\pgfusepath{stroke}%
\end{pgfscope}%
\begin{pgfscope}%
\pgfpathrectangle{\pgfqpoint{0.510806in}{0.456793in}}{\pgfqpoint{4.447524in}{2.201537in}}%
\pgfusepath{clip}%
\pgfsetrectcap%
\pgfsetroundjoin%
\pgfsetlinewidth{1.003750pt}%
\definecolor{currentstroke}{rgb}{0.000000,0.000000,0.000000}%
\pgfsetstrokecolor{currentstroke}%
\pgfsetdash{}{0pt}%
\pgfpathmoveto{\pgfqpoint{2.979182in}{1.110650in}}%
\pgfpathlineto{\pgfqpoint{3.112608in}{1.110650in}}%
\pgfusepath{stroke}%
\end{pgfscope}%
\begin{pgfscope}%
\pgfpathrectangle{\pgfqpoint{0.510806in}{0.456793in}}{\pgfqpoint{4.447524in}{2.201537in}}%
\pgfusepath{clip}%
\pgfsetrectcap%
\pgfsetroundjoin%
\pgfsetlinewidth{1.003750pt}%
\definecolor{currentstroke}{rgb}{0.000000,0.000000,0.000000}%
\pgfsetstrokecolor{currentstroke}%
\pgfsetdash{}{0pt}%
\pgfpathmoveto{\pgfqpoint{2.979182in}{2.005354in}}%
\pgfpathlineto{\pgfqpoint{3.112608in}{2.005354in}}%
\pgfusepath{stroke}%
\end{pgfscope}%
\begin{pgfscope}%
\pgfpathrectangle{\pgfqpoint{0.510806in}{0.456793in}}{\pgfqpoint{4.447524in}{2.201537in}}%
\pgfusepath{clip}%
\pgfsetbuttcap%
\pgfsetmiterjoin%
\definecolor{currentfill}{rgb}{0.552941,0.627451,0.796078}%
\pgfsetfillcolor{currentfill}%
\pgfsetlinewidth{1.003750pt}%
\definecolor{currentstroke}{rgb}{0.000000,0.000000,0.000000}%
\pgfsetstrokecolor{currentstroke}%
\pgfsetdash{}{0pt}%
\pgfpathmoveto{\pgfqpoint{4.024350in}{1.442862in}}%
\pgfpathlineto{\pgfqpoint{4.291201in}{1.442862in}}%
\pgfpathlineto{\pgfqpoint{4.291201in}{1.550911in}}%
\pgfpathlineto{\pgfqpoint{4.224489in}{1.552058in}}%
\pgfpathlineto{\pgfqpoint{4.291201in}{1.553204in}}%
\pgfpathlineto{\pgfqpoint{4.291201in}{1.673803in}}%
\pgfpathlineto{\pgfqpoint{4.024350in}{1.673803in}}%
\pgfpathlineto{\pgfqpoint{4.024350in}{1.553204in}}%
\pgfpathlineto{\pgfqpoint{4.091063in}{1.552058in}}%
\pgfpathlineto{\pgfqpoint{4.024350in}{1.550911in}}%
\pgfpathlineto{\pgfqpoint{4.024350in}{1.442862in}}%
\pgfpathlineto{\pgfqpoint{4.024350in}{1.442862in}}%
\pgfpathclose%
\pgfusepath{stroke,fill}%
\end{pgfscope}%
\begin{pgfscope}%
\pgfpathrectangle{\pgfqpoint{0.510806in}{0.456793in}}{\pgfqpoint{4.447524in}{2.201537in}}%
\pgfusepath{clip}%
\pgfsetrectcap%
\pgfsetroundjoin%
\pgfsetlinewidth{1.003750pt}%
\definecolor{currentstroke}{rgb}{0.000000,0.000000,0.000000}%
\pgfsetstrokecolor{currentstroke}%
\pgfsetdash{}{0pt}%
\pgfpathmoveto{\pgfqpoint{4.157776in}{1.442862in}}%
\pgfpathlineto{\pgfqpoint{4.157776in}{1.096560in}}%
\pgfusepath{stroke}%
\end{pgfscope}%
\begin{pgfscope}%
\pgfpathrectangle{\pgfqpoint{0.510806in}{0.456793in}}{\pgfqpoint{4.447524in}{2.201537in}}%
\pgfusepath{clip}%
\pgfsetrectcap%
\pgfsetroundjoin%
\pgfsetlinewidth{1.003750pt}%
\definecolor{currentstroke}{rgb}{0.000000,0.000000,0.000000}%
\pgfsetstrokecolor{currentstroke}%
\pgfsetdash{}{0pt}%
\pgfpathmoveto{\pgfqpoint{4.157776in}{1.673803in}}%
\pgfpathlineto{\pgfqpoint{4.157776in}{2.019884in}}%
\pgfusepath{stroke}%
\end{pgfscope}%
\begin{pgfscope}%
\pgfpathrectangle{\pgfqpoint{0.510806in}{0.456793in}}{\pgfqpoint{4.447524in}{2.201537in}}%
\pgfusepath{clip}%
\pgfsetrectcap%
\pgfsetroundjoin%
\pgfsetlinewidth{1.003750pt}%
\definecolor{currentstroke}{rgb}{0.000000,0.000000,0.000000}%
\pgfsetstrokecolor{currentstroke}%
\pgfsetdash{}{0pt}%
\pgfpathmoveto{\pgfqpoint{4.091063in}{1.096560in}}%
\pgfpathlineto{\pgfqpoint{4.224489in}{1.096560in}}%
\pgfusepath{stroke}%
\end{pgfscope}%
\begin{pgfscope}%
\pgfpathrectangle{\pgfqpoint{0.510806in}{0.456793in}}{\pgfqpoint{4.447524in}{2.201537in}}%
\pgfusepath{clip}%
\pgfsetrectcap%
\pgfsetroundjoin%
\pgfsetlinewidth{1.003750pt}%
\definecolor{currentstroke}{rgb}{0.000000,0.000000,0.000000}%
\pgfsetstrokecolor{currentstroke}%
\pgfsetdash{}{0pt}%
\pgfpathmoveto{\pgfqpoint{4.091063in}{2.019884in}}%
\pgfpathlineto{\pgfqpoint{4.224489in}{2.019884in}}%
\pgfusepath{stroke}%
\end{pgfscope}%
\begin{pgfscope}%
\pgfpathrectangle{\pgfqpoint{0.510806in}{0.456793in}}{\pgfqpoint{4.447524in}{2.201537in}}%
\pgfusepath{clip}%
\pgfsetrectcap%
\pgfsetroundjoin%
\pgfsetlinewidth{1.003750pt}%
\definecolor{currentstroke}{rgb}{0.000000,0.000000,0.000000}%
\pgfsetstrokecolor{currentstroke}%
\pgfsetdash{}{0pt}%
\pgfpathmoveto{\pgfqpoint{1.244648in}{1.554480in}}%
\pgfpathlineto{\pgfqpoint{1.378073in}{1.554480in}}%
\pgfusepath{stroke}%
\end{pgfscope}%
\begin{pgfscope}%
\pgfpathrectangle{\pgfqpoint{0.510806in}{0.456793in}}{\pgfqpoint{4.447524in}{2.201537in}}%
\pgfusepath{clip}%
\pgfsetrectcap%
\pgfsetroundjoin%
\pgfsetlinewidth{1.003750pt}%
\definecolor{currentstroke}{rgb}{0.000000,0.000000,0.000000}%
\pgfsetstrokecolor{currentstroke}%
\pgfsetdash{}{0pt}%
\pgfpathmoveto{\pgfqpoint{2.356529in}{1.550737in}}%
\pgfpathlineto{\pgfqpoint{2.489954in}{1.550737in}}%
\pgfusepath{stroke}%
\end{pgfscope}%
\begin{pgfscope}%
\pgfpathrectangle{\pgfqpoint{0.510806in}{0.456793in}}{\pgfqpoint{4.447524in}{2.201537in}}%
\pgfusepath{clip}%
\pgfsetrectcap%
\pgfsetroundjoin%
\pgfsetlinewidth{1.003750pt}%
\definecolor{currentstroke}{rgb}{0.000000,0.000000,0.000000}%
\pgfsetstrokecolor{currentstroke}%
\pgfsetdash{}{0pt}%
\pgfpathmoveto{\pgfqpoint{3.468410in}{1.537968in}}%
\pgfpathlineto{\pgfqpoint{3.601835in}{1.537968in}}%
\pgfusepath{stroke}%
\end{pgfscope}%
\begin{pgfscope}%
\pgfpathrectangle{\pgfqpoint{0.510806in}{0.456793in}}{\pgfqpoint{4.447524in}{2.201537in}}%
\pgfusepath{clip}%
\pgfsetrectcap%
\pgfsetroundjoin%
\pgfsetlinewidth{1.003750pt}%
\definecolor{currentstroke}{rgb}{0.000000,0.000000,0.000000}%
\pgfsetstrokecolor{currentstroke}%
\pgfsetdash{}{0pt}%
\pgfpathmoveto{\pgfqpoint{1.555974in}{1.546554in}}%
\pgfpathlineto{\pgfqpoint{1.689400in}{1.546554in}}%
\pgfusepath{stroke}%
\end{pgfscope}%
\begin{pgfscope}%
\pgfpathrectangle{\pgfqpoint{0.510806in}{0.456793in}}{\pgfqpoint{4.447524in}{2.201537in}}%
\pgfusepath{clip}%
\pgfsetrectcap%
\pgfsetroundjoin%
\pgfsetlinewidth{1.003750pt}%
\definecolor{currentstroke}{rgb}{0.000000,0.000000,0.000000}%
\pgfsetstrokecolor{currentstroke}%
\pgfsetdash{}{0pt}%
\pgfpathmoveto{\pgfqpoint{2.667855in}{1.554480in}}%
\pgfpathlineto{\pgfqpoint{2.801281in}{1.554480in}}%
\pgfusepath{stroke}%
\end{pgfscope}%
\begin{pgfscope}%
\pgfpathrectangle{\pgfqpoint{0.510806in}{0.456793in}}{\pgfqpoint{4.447524in}{2.201537in}}%
\pgfusepath{clip}%
\pgfsetrectcap%
\pgfsetroundjoin%
\pgfsetlinewidth{1.003750pt}%
\definecolor{currentstroke}{rgb}{0.000000,0.000000,0.000000}%
\pgfsetstrokecolor{currentstroke}%
\pgfsetdash{}{0pt}%
\pgfpathmoveto{\pgfqpoint{3.779736in}{1.550297in}}%
\pgfpathlineto{\pgfqpoint{3.913162in}{1.550297in}}%
\pgfusepath{stroke}%
\end{pgfscope}%
\begin{pgfscope}%
\pgfpathrectangle{\pgfqpoint{0.510806in}{0.456793in}}{\pgfqpoint{4.447524in}{2.201537in}}%
\pgfusepath{clip}%
\pgfsetrectcap%
\pgfsetroundjoin%
\pgfsetlinewidth{1.003750pt}%
\definecolor{currentstroke}{rgb}{0.000000,0.000000,0.000000}%
\pgfsetstrokecolor{currentstroke}%
\pgfsetdash{}{0pt}%
\pgfpathmoveto{\pgfqpoint{1.867301in}{1.551618in}}%
\pgfpathlineto{\pgfqpoint{2.000727in}{1.551618in}}%
\pgfusepath{stroke}%
\end{pgfscope}%
\begin{pgfscope}%
\pgfpathrectangle{\pgfqpoint{0.510806in}{0.456793in}}{\pgfqpoint{4.447524in}{2.201537in}}%
\pgfusepath{clip}%
\pgfsetrectcap%
\pgfsetroundjoin%
\pgfsetlinewidth{1.003750pt}%
\definecolor{currentstroke}{rgb}{0.000000,0.000000,0.000000}%
\pgfsetstrokecolor{currentstroke}%
\pgfsetdash{}{0pt}%
\pgfpathmoveto{\pgfqpoint{2.979182in}{1.554920in}}%
\pgfpathlineto{\pgfqpoint{3.112608in}{1.554920in}}%
\pgfusepath{stroke}%
\end{pgfscope}%
\begin{pgfscope}%
\pgfpathrectangle{\pgfqpoint{0.510806in}{0.456793in}}{\pgfqpoint{4.447524in}{2.201537in}}%
\pgfusepath{clip}%
\pgfsetrectcap%
\pgfsetroundjoin%
\pgfsetlinewidth{1.003750pt}%
\definecolor{currentstroke}{rgb}{0.000000,0.000000,0.000000}%
\pgfsetstrokecolor{currentstroke}%
\pgfsetdash{}{0pt}%
\pgfpathmoveto{\pgfqpoint{4.091063in}{1.552058in}}%
\pgfpathlineto{\pgfqpoint{4.224489in}{1.552058in}}%
\pgfusepath{stroke}%
\end{pgfscope}%
\begin{pgfscope}%
\pgfsetrectcap%
\pgfsetmiterjoin%
\pgfsetlinewidth{0.803000pt}%
\definecolor{currentstroke}{rgb}{0.000000,0.000000,0.000000}%
\pgfsetstrokecolor{currentstroke}%
\pgfsetdash{}{0pt}%
\pgfpathmoveto{\pgfqpoint{0.510806in}{0.456793in}}%
\pgfpathlineto{\pgfqpoint{0.510806in}{2.658330in}}%
\pgfusepath{stroke}%
\end{pgfscope}%
\begin{pgfscope}%
\pgfsetrectcap%
\pgfsetmiterjoin%
\pgfsetlinewidth{0.803000pt}%
\definecolor{currentstroke}{rgb}{0.000000,0.000000,0.000000}%
\pgfsetstrokecolor{currentstroke}%
\pgfsetdash{}{0pt}%
\pgfpathmoveto{\pgfqpoint{4.958330in}{0.456793in}}%
\pgfpathlineto{\pgfqpoint{4.958330in}{2.658330in}}%
\pgfusepath{stroke}%
\end{pgfscope}%
\begin{pgfscope}%
\pgfsetrectcap%
\pgfsetmiterjoin%
\pgfsetlinewidth{0.803000pt}%
\definecolor{currentstroke}{rgb}{0.000000,0.000000,0.000000}%
\pgfsetstrokecolor{currentstroke}%
\pgfsetdash{}{0pt}%
\pgfpathmoveto{\pgfqpoint{0.510806in}{0.456793in}}%
\pgfpathlineto{\pgfqpoint{4.958330in}{0.456793in}}%
\pgfusepath{stroke}%
\end{pgfscope}%
\begin{pgfscope}%
\pgfsetrectcap%
\pgfsetmiterjoin%
\pgfsetlinewidth{0.803000pt}%
\definecolor{currentstroke}{rgb}{0.000000,0.000000,0.000000}%
\pgfsetstrokecolor{currentstroke}%
\pgfsetdash{}{0pt}%
\pgfpathmoveto{\pgfqpoint{0.510806in}{2.658330in}}%
\pgfpathlineto{\pgfqpoint{4.958330in}{2.658330in}}%
\pgfusepath{stroke}%
\end{pgfscope}%
\begin{pgfscope}%
\pgfsetbuttcap%
\pgfsetmiterjoin%
\definecolor{currentfill}{rgb}{1.000000,1.000000,1.000000}%
\pgfsetfillcolor{currentfill}%
\pgfsetfillopacity{0.800000}%
\pgfsetlinewidth{1.003750pt}%
\definecolor{currentstroke}{rgb}{0.800000,0.800000,0.800000}%
\pgfsetstrokecolor{currentstroke}%
\pgfsetstrokeopacity{0.800000}%
\pgfsetdash{}{0pt}%
\pgfpathmoveto{\pgfqpoint{1.349916in}{2.353546in}}%
\pgfpathlineto{\pgfqpoint{4.888886in}{2.353546in}}%
\pgfpathlineto{\pgfqpoint{4.888886in}{2.588886in}}%
\pgfpathlineto{\pgfqpoint{1.349916in}{2.588886in}}%
\pgfpathlineto{\pgfqpoint{1.349916in}{2.353546in}}%
\pgfpathclose%
\pgfusepath{stroke,fill}%
\end{pgfscope}%
\begin{pgfscope}%
\pgfsetbuttcap%
\pgfsetmiterjoin%
\definecolor{currentfill}{rgb}{0.400000,0.760784,0.647059}%
\pgfsetfillcolor{currentfill}%
\pgfsetlinewidth{1.003750pt}%
\definecolor{currentstroke}{rgb}{0.000000,0.000000,0.000000}%
\pgfsetstrokecolor{currentstroke}%
\pgfsetdash{}{0pt}%
\pgfpathmoveto{\pgfqpoint{1.405472in}{2.436108in}}%
\pgfpathlineto{\pgfqpoint{1.683249in}{2.436108in}}%
\pgfpathlineto{\pgfqpoint{1.683249in}{2.533330in}}%
\pgfpathlineto{\pgfqpoint{1.405472in}{2.533330in}}%
\pgfpathlineto{\pgfqpoint{1.405472in}{2.436108in}}%
\pgfpathclose%
\pgfusepath{stroke,fill}%
\end{pgfscope}%
\begin{pgfscope}%
\definecolor{textcolor}{rgb}{0.000000,0.000000,0.000000}%
\pgfsetstrokecolor{textcolor}%
\pgfsetfillcolor{textcolor}%
\pgftext[x=1.794361in,y=2.436108in,left,base]{\color{textcolor}\rmfamily\fontsize{10.000000}{12.000000}\selectfont Bare-metal}%
\end{pgfscope}%
\begin{pgfscope}%
\pgfsetbuttcap%
\pgfsetmiterjoin%
\definecolor{currentfill}{rgb}{0.988235,0.552941,0.384314}%
\pgfsetfillcolor{currentfill}%
\pgfsetlinewidth{1.003750pt}%
\definecolor{currentstroke}{rgb}{0.000000,0.000000,0.000000}%
\pgfsetstrokecolor{currentstroke}%
\pgfsetdash{}{0pt}%
\pgfpathmoveto{\pgfqpoint{2.741893in}{2.436108in}}%
\pgfpathlineto{\pgfqpoint{3.019671in}{2.436108in}}%
\pgfpathlineto{\pgfqpoint{3.019671in}{2.533330in}}%
\pgfpathlineto{\pgfqpoint{2.741893in}{2.533330in}}%
\pgfpathlineto{\pgfqpoint{2.741893in}{2.436108in}}%
\pgfpathclose%
\pgfusepath{stroke,fill}%
\end{pgfscope}%
\begin{pgfscope}%
\definecolor{textcolor}{rgb}{0.000000,0.000000,0.000000}%
\pgfsetstrokecolor{textcolor}%
\pgfsetfillcolor{textcolor}%
\pgftext[x=3.130782in,y=2.436108in,left,base]{\color{textcolor}\rmfamily\fontsize{10.000000}{12.000000}\selectfont TSN-CNI}%
\end{pgfscope}%
\begin{pgfscope}%
\pgfsetbuttcap%
\pgfsetmiterjoin%
\definecolor{currentfill}{rgb}{0.552941,0.627451,0.796078}%
\pgfsetfillcolor{currentfill}%
\pgfsetlinewidth{1.003750pt}%
\definecolor{currentstroke}{rgb}{0.000000,0.000000,0.000000}%
\pgfsetstrokecolor{currentstroke}%
\pgfsetdash{}{0pt}%
\pgfpathmoveto{\pgfqpoint{3.991122in}{2.436108in}}%
\pgfpathlineto{\pgfqpoint{4.268900in}{2.436108in}}%
\pgfpathlineto{\pgfqpoint{4.268900in}{2.533330in}}%
\pgfpathlineto{\pgfqpoint{3.991122in}{2.533330in}}%
\pgfpathlineto{\pgfqpoint{3.991122in}{2.436108in}}%
\pgfpathclose%
\pgfusepath{stroke,fill}%
\end{pgfscope}%
\begin{pgfscope}%
\definecolor{textcolor}{rgb}{0.000000,0.000000,0.000000}%
\pgfsetstrokecolor{textcolor}%
\pgfsetfillcolor{textcolor}%
\pgftext[x=4.380011in,y=2.436108in,left,base]{\color{textcolor}\rmfamily\fontsize{10.000000}{12.000000}\selectfont Flannel}%
\end{pgfscope}%
\end{pgfpicture}%
\makeatother%
\endgroup%

%% file: cdf.pgf
%% Creator: Matplotlib, PGF backend
%%
%% To include the figure in your LaTeX document, write
%%   \input{<filename>.pgf}
%%
%% Make sure the required packages are loaded in your preamble
%%   \usepackage{pgf}
%%
%% Also ensure that all the required font packages are loaded; for instance,
%% the lmodern package is sometimes necessary when using math font.
%%   \usepackage{lmodern}
%%
%% Figures using additional raster images can only be included by \input if
%% they are in the same directory as the main LaTeX file. For loading figures
%% from other directories you can use the `import` package
%%   \usepackage{import}
%%
%% and then include the figures with
%%   \import{<path to file>}{<filename>.pgf}
%%
%% Matplotlib used the following preamble
%%   
%%   \makeatletter\@ifpackageloaded{underscore}{}{\usepackage[strings]{underscore}}\makeatother
%%
\begingroup%
\makeatletter%
\begin{pgfpicture}%
\pgfpathrectangle{\pgfpointorigin}{\pgfqpoint{5.000000in}{2.700000in}}%
\pgfusepath{use as bounding box, clip}%
\begin{pgfscope}%
\pgfsetbuttcap%
\pgfsetmiterjoin%
\definecolor{currentfill}{rgb}{1.000000,1.000000,1.000000}%
\pgfsetfillcolor{currentfill}%
\pgfsetlinewidth{0.000000pt}%
\definecolor{currentstroke}{rgb}{1.000000,1.000000,1.000000}%
\pgfsetstrokecolor{currentstroke}%
\pgfsetdash{}{0pt}%
\pgfpathmoveto{\pgfqpoint{0.000000in}{0.000000in}}%
\pgfpathlineto{\pgfqpoint{5.000000in}{0.000000in}}%
\pgfpathlineto{\pgfqpoint{5.000000in}{2.700000in}}%
\pgfpathlineto{\pgfqpoint{0.000000in}{2.700000in}}%
\pgfpathlineto{\pgfqpoint{0.000000in}{0.000000in}}%
\pgfpathclose%
\pgfusepath{fill}%
\end{pgfscope}%
\begin{pgfscope}%
\pgfsetbuttcap%
\pgfsetmiterjoin%
\definecolor{currentfill}{rgb}{1.000000,1.000000,1.000000}%
\pgfsetfillcolor{currentfill}%
\pgfsetlinewidth{0.000000pt}%
\definecolor{currentstroke}{rgb}{0.000000,0.000000,0.000000}%
\pgfsetstrokecolor{currentstroke}%
\pgfsetstrokeopacity{0.000000}%
\pgfsetdash{}{0pt}%
\pgfpathmoveto{\pgfqpoint{0.495374in}{0.456793in}}%
\pgfpathlineto{\pgfqpoint{4.888885in}{0.456793in}}%
\pgfpathlineto{\pgfqpoint{4.888885in}{2.658330in}}%
\pgfpathlineto{\pgfqpoint{0.495374in}{2.658330in}}%
\pgfpathlineto{\pgfqpoint{0.495374in}{0.456793in}}%
\pgfpathclose%
\pgfusepath{fill}%
\end{pgfscope}%
\begin{pgfscope}%
\pgfpathrectangle{\pgfqpoint{0.495374in}{0.456793in}}{\pgfqpoint{4.393511in}{2.201537in}}%
\pgfusepath{clip}%
\pgfsetbuttcap%
\pgfsetroundjoin%
\pgfsetlinewidth{0.803000pt}%
\definecolor{currentstroke}{rgb}{0.690196,0.690196,0.690196}%
\pgfsetstrokecolor{currentstroke}%
\pgfsetdash{{0.800000pt}{1.320000pt}}{0.000000pt}%
\pgfpathmoveto{\pgfqpoint{0.495374in}{0.456793in}}%
\pgfpathlineto{\pgfqpoint{0.495374in}{2.658330in}}%
\pgfusepath{stroke}%
\end{pgfscope}%
\begin{pgfscope}%
\pgfsetbuttcap%
\pgfsetroundjoin%
\definecolor{currentfill}{rgb}{0.000000,0.000000,0.000000}%
\pgfsetfillcolor{currentfill}%
\pgfsetlinewidth{0.803000pt}%
\definecolor{currentstroke}{rgb}{0.000000,0.000000,0.000000}%
\pgfsetstrokecolor{currentstroke}%
\pgfsetdash{}{0pt}%
\pgfsys@defobject{currentmarker}{\pgfqpoint{0.000000in}{-0.048611in}}{\pgfqpoint{0.000000in}{0.000000in}}{%
\pgfpathmoveto{\pgfqpoint{0.000000in}{0.000000in}}%
\pgfpathlineto{\pgfqpoint{0.000000in}{-0.048611in}}%
\pgfusepath{stroke,fill}%
}%
\begin{pgfscope}%
\pgfsys@transformshift{0.495374in}{0.456793in}%
\pgfsys@useobject{currentmarker}{}%
\end{pgfscope}%
\end{pgfscope}%
\begin{pgfscope}%
\definecolor{textcolor}{rgb}{0.000000,0.000000,0.000000}%
\pgfsetstrokecolor{textcolor}%
\pgfsetfillcolor{textcolor}%
\pgftext[x=0.495374in,y=0.359571in,,top]{\color{textcolor}\rmfamily\fontsize{10.000000}{12.000000}\selectfont \(\displaystyle {0}\)}%
\end{pgfscope}%
\begin{pgfscope}%
\pgfpathrectangle{\pgfqpoint{0.495374in}{0.456793in}}{\pgfqpoint{4.393511in}{2.201537in}}%
\pgfusepath{clip}%
\pgfsetbuttcap%
\pgfsetroundjoin%
\pgfsetlinewidth{0.803000pt}%
\definecolor{currentstroke}{rgb}{0.690196,0.690196,0.690196}%
\pgfsetstrokecolor{currentstroke}%
\pgfsetdash{{0.800000pt}{1.320000pt}}{0.000000pt}%
\pgfpathmoveto{\pgfqpoint{1.374076in}{0.456793in}}%
\pgfpathlineto{\pgfqpoint{1.374076in}{2.658330in}}%
\pgfusepath{stroke}%
\end{pgfscope}%
\begin{pgfscope}%
\pgfsetbuttcap%
\pgfsetroundjoin%
\definecolor{currentfill}{rgb}{0.000000,0.000000,0.000000}%
\pgfsetfillcolor{currentfill}%
\pgfsetlinewidth{0.803000pt}%
\definecolor{currentstroke}{rgb}{0.000000,0.000000,0.000000}%
\pgfsetstrokecolor{currentstroke}%
\pgfsetdash{}{0pt}%
\pgfsys@defobject{currentmarker}{\pgfqpoint{0.000000in}{-0.048611in}}{\pgfqpoint{0.000000in}{0.000000in}}{%
\pgfpathmoveto{\pgfqpoint{0.000000in}{0.000000in}}%
\pgfpathlineto{\pgfqpoint{0.000000in}{-0.048611in}}%
\pgfusepath{stroke,fill}%
}%
\begin{pgfscope}%
\pgfsys@transformshift{1.374076in}{0.456793in}%
\pgfsys@useobject{currentmarker}{}%
\end{pgfscope}%
\end{pgfscope}%
\begin{pgfscope}%
\definecolor{textcolor}{rgb}{0.000000,0.000000,0.000000}%
\pgfsetstrokecolor{textcolor}%
\pgfsetfillcolor{textcolor}%
\pgftext[x=1.374076in,y=0.359571in,,top]{\color{textcolor}\rmfamily\fontsize{10.000000}{12.000000}\selectfont \(\displaystyle {10}\)}%
\end{pgfscope}%
\begin{pgfscope}%
\pgfpathrectangle{\pgfqpoint{0.495374in}{0.456793in}}{\pgfqpoint{4.393511in}{2.201537in}}%
\pgfusepath{clip}%
\pgfsetbuttcap%
\pgfsetroundjoin%
\pgfsetlinewidth{0.803000pt}%
\definecolor{currentstroke}{rgb}{0.690196,0.690196,0.690196}%
\pgfsetstrokecolor{currentstroke}%
\pgfsetdash{{0.800000pt}{1.320000pt}}{0.000000pt}%
\pgfpathmoveto{\pgfqpoint{2.252779in}{0.456793in}}%
\pgfpathlineto{\pgfqpoint{2.252779in}{2.658330in}}%
\pgfusepath{stroke}%
\end{pgfscope}%
\begin{pgfscope}%
\pgfsetbuttcap%
\pgfsetroundjoin%
\definecolor{currentfill}{rgb}{0.000000,0.000000,0.000000}%
\pgfsetfillcolor{currentfill}%
\pgfsetlinewidth{0.803000pt}%
\definecolor{currentstroke}{rgb}{0.000000,0.000000,0.000000}%
\pgfsetstrokecolor{currentstroke}%
\pgfsetdash{}{0pt}%
\pgfsys@defobject{currentmarker}{\pgfqpoint{0.000000in}{-0.048611in}}{\pgfqpoint{0.000000in}{0.000000in}}{%
\pgfpathmoveto{\pgfqpoint{0.000000in}{0.000000in}}%
\pgfpathlineto{\pgfqpoint{0.000000in}{-0.048611in}}%
\pgfusepath{stroke,fill}%
}%
\begin{pgfscope}%
\pgfsys@transformshift{2.252779in}{0.456793in}%
\pgfsys@useobject{currentmarker}{}%
\end{pgfscope}%
\end{pgfscope}%
\begin{pgfscope}%
\definecolor{textcolor}{rgb}{0.000000,0.000000,0.000000}%
\pgfsetstrokecolor{textcolor}%
\pgfsetfillcolor{textcolor}%
\pgftext[x=2.252779in,y=0.359571in,,top]{\color{textcolor}\rmfamily\fontsize{10.000000}{12.000000}\selectfont \(\displaystyle {20}\)}%
\end{pgfscope}%
\begin{pgfscope}%
\pgfpathrectangle{\pgfqpoint{0.495374in}{0.456793in}}{\pgfqpoint{4.393511in}{2.201537in}}%
\pgfusepath{clip}%
\pgfsetbuttcap%
\pgfsetroundjoin%
\pgfsetlinewidth{0.803000pt}%
\definecolor{currentstroke}{rgb}{0.690196,0.690196,0.690196}%
\pgfsetstrokecolor{currentstroke}%
\pgfsetdash{{0.800000pt}{1.320000pt}}{0.000000pt}%
\pgfpathmoveto{\pgfqpoint{3.131481in}{0.456793in}}%
\pgfpathlineto{\pgfqpoint{3.131481in}{2.658330in}}%
\pgfusepath{stroke}%
\end{pgfscope}%
\begin{pgfscope}%
\pgfsetbuttcap%
\pgfsetroundjoin%
\definecolor{currentfill}{rgb}{0.000000,0.000000,0.000000}%
\pgfsetfillcolor{currentfill}%
\pgfsetlinewidth{0.803000pt}%
\definecolor{currentstroke}{rgb}{0.000000,0.000000,0.000000}%
\pgfsetstrokecolor{currentstroke}%
\pgfsetdash{}{0pt}%
\pgfsys@defobject{currentmarker}{\pgfqpoint{0.000000in}{-0.048611in}}{\pgfqpoint{0.000000in}{0.000000in}}{%
\pgfpathmoveto{\pgfqpoint{0.000000in}{0.000000in}}%
\pgfpathlineto{\pgfqpoint{0.000000in}{-0.048611in}}%
\pgfusepath{stroke,fill}%
}%
\begin{pgfscope}%
\pgfsys@transformshift{3.131481in}{0.456793in}%
\pgfsys@useobject{currentmarker}{}%
\end{pgfscope}%
\end{pgfscope}%
\begin{pgfscope}%
\definecolor{textcolor}{rgb}{0.000000,0.000000,0.000000}%
\pgfsetstrokecolor{textcolor}%
\pgfsetfillcolor{textcolor}%
\pgftext[x=3.131481in,y=0.359571in,,top]{\color{textcolor}\rmfamily\fontsize{10.000000}{12.000000}\selectfont \(\displaystyle {30}\)}%
\end{pgfscope}%
\begin{pgfscope}%
\pgfpathrectangle{\pgfqpoint{0.495374in}{0.456793in}}{\pgfqpoint{4.393511in}{2.201537in}}%
\pgfusepath{clip}%
\pgfsetbuttcap%
\pgfsetroundjoin%
\pgfsetlinewidth{0.803000pt}%
\definecolor{currentstroke}{rgb}{0.690196,0.690196,0.690196}%
\pgfsetstrokecolor{currentstroke}%
\pgfsetdash{{0.800000pt}{1.320000pt}}{0.000000pt}%
\pgfpathmoveto{\pgfqpoint{4.010183in}{0.456793in}}%
\pgfpathlineto{\pgfqpoint{4.010183in}{2.658330in}}%
\pgfusepath{stroke}%
\end{pgfscope}%
\begin{pgfscope}%
\pgfsetbuttcap%
\pgfsetroundjoin%
\definecolor{currentfill}{rgb}{0.000000,0.000000,0.000000}%
\pgfsetfillcolor{currentfill}%
\pgfsetlinewidth{0.803000pt}%
\definecolor{currentstroke}{rgb}{0.000000,0.000000,0.000000}%
\pgfsetstrokecolor{currentstroke}%
\pgfsetdash{}{0pt}%
\pgfsys@defobject{currentmarker}{\pgfqpoint{0.000000in}{-0.048611in}}{\pgfqpoint{0.000000in}{0.000000in}}{%
\pgfpathmoveto{\pgfqpoint{0.000000in}{0.000000in}}%
\pgfpathlineto{\pgfqpoint{0.000000in}{-0.048611in}}%
\pgfusepath{stroke,fill}%
}%
\begin{pgfscope}%
\pgfsys@transformshift{4.010183in}{0.456793in}%
\pgfsys@useobject{currentmarker}{}%
\end{pgfscope}%
\end{pgfscope}%
\begin{pgfscope}%
\definecolor{textcolor}{rgb}{0.000000,0.000000,0.000000}%
\pgfsetstrokecolor{textcolor}%
\pgfsetfillcolor{textcolor}%
\pgftext[x=4.010183in,y=0.359571in,,top]{\color{textcolor}\rmfamily\fontsize{10.000000}{12.000000}\selectfont \(\displaystyle {40}\)}%
\end{pgfscope}%
\begin{pgfscope}%
\pgfpathrectangle{\pgfqpoint{0.495374in}{0.456793in}}{\pgfqpoint{4.393511in}{2.201537in}}%
\pgfusepath{clip}%
\pgfsetbuttcap%
\pgfsetroundjoin%
\pgfsetlinewidth{0.803000pt}%
\definecolor{currentstroke}{rgb}{0.690196,0.690196,0.690196}%
\pgfsetstrokecolor{currentstroke}%
\pgfsetdash{{0.800000pt}{1.320000pt}}{0.000000pt}%
\pgfpathmoveto{\pgfqpoint{4.888885in}{0.456793in}}%
\pgfpathlineto{\pgfqpoint{4.888885in}{2.658330in}}%
\pgfusepath{stroke}%
\end{pgfscope}%
\begin{pgfscope}%
\pgfsetbuttcap%
\pgfsetroundjoin%
\definecolor{currentfill}{rgb}{0.000000,0.000000,0.000000}%
\pgfsetfillcolor{currentfill}%
\pgfsetlinewidth{0.803000pt}%
\definecolor{currentstroke}{rgb}{0.000000,0.000000,0.000000}%
\pgfsetstrokecolor{currentstroke}%
\pgfsetdash{}{0pt}%
\pgfsys@defobject{currentmarker}{\pgfqpoint{0.000000in}{-0.048611in}}{\pgfqpoint{0.000000in}{0.000000in}}{%
\pgfpathmoveto{\pgfqpoint{0.000000in}{0.000000in}}%
\pgfpathlineto{\pgfqpoint{0.000000in}{-0.048611in}}%
\pgfusepath{stroke,fill}%
}%
\begin{pgfscope}%
\pgfsys@transformshift{4.888885in}{0.456793in}%
\pgfsys@useobject{currentmarker}{}%
\end{pgfscope}%
\end{pgfscope}%
\begin{pgfscope}%
\definecolor{textcolor}{rgb}{0.000000,0.000000,0.000000}%
\pgfsetstrokecolor{textcolor}%
\pgfsetfillcolor{textcolor}%
\pgftext[x=4.888885in,y=0.359571in,,top]{\color{textcolor}\rmfamily\fontsize{10.000000}{12.000000}\selectfont \(\displaystyle {50}\)}%
\end{pgfscope}%
\begin{pgfscope}%
\definecolor{textcolor}{rgb}{0.000000,0.000000,0.000000}%
\pgfsetstrokecolor{textcolor}%
\pgfsetfillcolor{textcolor}%
\pgftext[x=2.692130in,y=0.180559in,,top]{\color{textcolor}\rmfamily\fontsize{10.000000}{12.000000}\selectfont Latency (\(\displaystyle \mu\)s)}%
\end{pgfscope}%
\begin{pgfscope}%
\pgfpathrectangle{\pgfqpoint{0.495374in}{0.456793in}}{\pgfqpoint{4.393511in}{2.201537in}}%
\pgfusepath{clip}%
\pgfsetbuttcap%
\pgfsetroundjoin%
\pgfsetlinewidth{0.803000pt}%
\definecolor{currentstroke}{rgb}{0.690196,0.690196,0.690196}%
\pgfsetstrokecolor{currentstroke}%
\pgfsetdash{{0.800000pt}{1.320000pt}}{0.000000pt}%
\pgfpathmoveto{\pgfqpoint{0.495374in}{0.556863in}}%
\pgfpathlineto{\pgfqpoint{4.888885in}{0.556863in}}%
\pgfusepath{stroke}%
\end{pgfscope}%
\begin{pgfscope}%
\pgfsetbuttcap%
\pgfsetroundjoin%
\definecolor{currentfill}{rgb}{0.000000,0.000000,0.000000}%
\pgfsetfillcolor{currentfill}%
\pgfsetlinewidth{0.803000pt}%
\definecolor{currentstroke}{rgb}{0.000000,0.000000,0.000000}%
\pgfsetstrokecolor{currentstroke}%
\pgfsetdash{}{0pt}%
\pgfsys@defobject{currentmarker}{\pgfqpoint{-0.048611in}{0.000000in}}{\pgfqpoint{-0.000000in}{0.000000in}}{%
\pgfpathmoveto{\pgfqpoint{-0.000000in}{0.000000in}}%
\pgfpathlineto{\pgfqpoint{-0.048611in}{0.000000in}}%
\pgfusepath{stroke,fill}%
}%
\begin{pgfscope}%
\pgfsys@transformshift{0.495374in}{0.556863in}%
\pgfsys@useobject{currentmarker}{}%
\end{pgfscope}%
\end{pgfscope}%
\begin{pgfscope}%
\definecolor{textcolor}{rgb}{0.000000,0.000000,0.000000}%
\pgfsetstrokecolor{textcolor}%
\pgfsetfillcolor{textcolor}%
\pgftext[x=0.220682in, y=0.508638in, left, base]{\color{textcolor}\rmfamily\fontsize{10.000000}{12.000000}\selectfont \(\displaystyle {0.0}\)}%
\end{pgfscope}%
\begin{pgfscope}%
\pgfpathrectangle{\pgfqpoint{0.495374in}{0.456793in}}{\pgfqpoint{4.393511in}{2.201537in}}%
\pgfusepath{clip}%
\pgfsetbuttcap%
\pgfsetroundjoin%
\pgfsetlinewidth{0.803000pt}%
\definecolor{currentstroke}{rgb}{0.690196,0.690196,0.690196}%
\pgfsetstrokecolor{currentstroke}%
\pgfsetdash{{0.800000pt}{1.320000pt}}{0.000000pt}%
\pgfpathmoveto{\pgfqpoint{0.495374in}{0.957143in}}%
\pgfpathlineto{\pgfqpoint{4.888885in}{0.957143in}}%
\pgfusepath{stroke}%
\end{pgfscope}%
\begin{pgfscope}%
\pgfsetbuttcap%
\pgfsetroundjoin%
\definecolor{currentfill}{rgb}{0.000000,0.000000,0.000000}%
\pgfsetfillcolor{currentfill}%
\pgfsetlinewidth{0.803000pt}%
\definecolor{currentstroke}{rgb}{0.000000,0.000000,0.000000}%
\pgfsetstrokecolor{currentstroke}%
\pgfsetdash{}{0pt}%
\pgfsys@defobject{currentmarker}{\pgfqpoint{-0.048611in}{0.000000in}}{\pgfqpoint{-0.000000in}{0.000000in}}{%
\pgfpathmoveto{\pgfqpoint{-0.000000in}{0.000000in}}%
\pgfpathlineto{\pgfqpoint{-0.048611in}{0.000000in}}%
\pgfusepath{stroke,fill}%
}%
\begin{pgfscope}%
\pgfsys@transformshift{0.495374in}{0.957143in}%
\pgfsys@useobject{currentmarker}{}%
\end{pgfscope}%
\end{pgfscope}%
\begin{pgfscope}%
\definecolor{textcolor}{rgb}{0.000000,0.000000,0.000000}%
\pgfsetstrokecolor{textcolor}%
\pgfsetfillcolor{textcolor}%
\pgftext[x=0.220682in, y=0.908917in, left, base]{\color{textcolor}\rmfamily\fontsize{10.000000}{12.000000}\selectfont \(\displaystyle {0.2}\)}%
\end{pgfscope}%
\begin{pgfscope}%
\pgfpathrectangle{\pgfqpoint{0.495374in}{0.456793in}}{\pgfqpoint{4.393511in}{2.201537in}}%
\pgfusepath{clip}%
\pgfsetbuttcap%
\pgfsetroundjoin%
\pgfsetlinewidth{0.803000pt}%
\definecolor{currentstroke}{rgb}{0.690196,0.690196,0.690196}%
\pgfsetstrokecolor{currentstroke}%
\pgfsetdash{{0.800000pt}{1.320000pt}}{0.000000pt}%
\pgfpathmoveto{\pgfqpoint{0.495374in}{1.357422in}}%
\pgfpathlineto{\pgfqpoint{4.888885in}{1.357422in}}%
\pgfusepath{stroke}%
\end{pgfscope}%
\begin{pgfscope}%
\pgfsetbuttcap%
\pgfsetroundjoin%
\definecolor{currentfill}{rgb}{0.000000,0.000000,0.000000}%
\pgfsetfillcolor{currentfill}%
\pgfsetlinewidth{0.803000pt}%
\definecolor{currentstroke}{rgb}{0.000000,0.000000,0.000000}%
\pgfsetstrokecolor{currentstroke}%
\pgfsetdash{}{0pt}%
\pgfsys@defobject{currentmarker}{\pgfqpoint{-0.048611in}{0.000000in}}{\pgfqpoint{-0.000000in}{0.000000in}}{%
\pgfpathmoveto{\pgfqpoint{-0.000000in}{0.000000in}}%
\pgfpathlineto{\pgfqpoint{-0.048611in}{0.000000in}}%
\pgfusepath{stroke,fill}%
}%
\begin{pgfscope}%
\pgfsys@transformshift{0.495374in}{1.357422in}%
\pgfsys@useobject{currentmarker}{}%
\end{pgfscope}%
\end{pgfscope}%
\begin{pgfscope}%
\definecolor{textcolor}{rgb}{0.000000,0.000000,0.000000}%
\pgfsetstrokecolor{textcolor}%
\pgfsetfillcolor{textcolor}%
\pgftext[x=0.220682in, y=1.309197in, left, base]{\color{textcolor}\rmfamily\fontsize{10.000000}{12.000000}\selectfont \(\displaystyle {0.4}\)}%
\end{pgfscope}%
\begin{pgfscope}%
\pgfpathrectangle{\pgfqpoint{0.495374in}{0.456793in}}{\pgfqpoint{4.393511in}{2.201537in}}%
\pgfusepath{clip}%
\pgfsetbuttcap%
\pgfsetroundjoin%
\pgfsetlinewidth{0.803000pt}%
\definecolor{currentstroke}{rgb}{0.690196,0.690196,0.690196}%
\pgfsetstrokecolor{currentstroke}%
\pgfsetdash{{0.800000pt}{1.320000pt}}{0.000000pt}%
\pgfpathmoveto{\pgfqpoint{0.495374in}{1.757701in}}%
\pgfpathlineto{\pgfqpoint{4.888885in}{1.757701in}}%
\pgfusepath{stroke}%
\end{pgfscope}%
\begin{pgfscope}%
\pgfsetbuttcap%
\pgfsetroundjoin%
\definecolor{currentfill}{rgb}{0.000000,0.000000,0.000000}%
\pgfsetfillcolor{currentfill}%
\pgfsetlinewidth{0.803000pt}%
\definecolor{currentstroke}{rgb}{0.000000,0.000000,0.000000}%
\pgfsetstrokecolor{currentstroke}%
\pgfsetdash{}{0pt}%
\pgfsys@defobject{currentmarker}{\pgfqpoint{-0.048611in}{0.000000in}}{\pgfqpoint{-0.000000in}{0.000000in}}{%
\pgfpathmoveto{\pgfqpoint{-0.000000in}{0.000000in}}%
\pgfpathlineto{\pgfqpoint{-0.048611in}{0.000000in}}%
\pgfusepath{stroke,fill}%
}%
\begin{pgfscope}%
\pgfsys@transformshift{0.495374in}{1.757701in}%
\pgfsys@useobject{currentmarker}{}%
\end{pgfscope}%
\end{pgfscope}%
\begin{pgfscope}%
\definecolor{textcolor}{rgb}{0.000000,0.000000,0.000000}%
\pgfsetstrokecolor{textcolor}%
\pgfsetfillcolor{textcolor}%
\pgftext[x=0.220682in, y=1.709476in, left, base]{\color{textcolor}\rmfamily\fontsize{10.000000}{12.000000}\selectfont \(\displaystyle {0.6}\)}%
\end{pgfscope}%
\begin{pgfscope}%
\pgfpathrectangle{\pgfqpoint{0.495374in}{0.456793in}}{\pgfqpoint{4.393511in}{2.201537in}}%
\pgfusepath{clip}%
\pgfsetbuttcap%
\pgfsetroundjoin%
\pgfsetlinewidth{0.803000pt}%
\definecolor{currentstroke}{rgb}{0.690196,0.690196,0.690196}%
\pgfsetstrokecolor{currentstroke}%
\pgfsetdash{{0.800000pt}{1.320000pt}}{0.000000pt}%
\pgfpathmoveto{\pgfqpoint{0.495374in}{2.157981in}}%
\pgfpathlineto{\pgfqpoint{4.888885in}{2.157981in}}%
\pgfusepath{stroke}%
\end{pgfscope}%
\begin{pgfscope}%
\pgfsetbuttcap%
\pgfsetroundjoin%
\definecolor{currentfill}{rgb}{0.000000,0.000000,0.000000}%
\pgfsetfillcolor{currentfill}%
\pgfsetlinewidth{0.803000pt}%
\definecolor{currentstroke}{rgb}{0.000000,0.000000,0.000000}%
\pgfsetstrokecolor{currentstroke}%
\pgfsetdash{}{0pt}%
\pgfsys@defobject{currentmarker}{\pgfqpoint{-0.048611in}{0.000000in}}{\pgfqpoint{-0.000000in}{0.000000in}}{%
\pgfpathmoveto{\pgfqpoint{-0.000000in}{0.000000in}}%
\pgfpathlineto{\pgfqpoint{-0.048611in}{0.000000in}}%
\pgfusepath{stroke,fill}%
}%
\begin{pgfscope}%
\pgfsys@transformshift{0.495374in}{2.157981in}%
\pgfsys@useobject{currentmarker}{}%
\end{pgfscope}%
\end{pgfscope}%
\begin{pgfscope}%
\definecolor{textcolor}{rgb}{0.000000,0.000000,0.000000}%
\pgfsetstrokecolor{textcolor}%
\pgfsetfillcolor{textcolor}%
\pgftext[x=0.220682in, y=2.109755in, left, base]{\color{textcolor}\rmfamily\fontsize{10.000000}{12.000000}\selectfont \(\displaystyle {0.8}\)}%
\end{pgfscope}%
\begin{pgfscope}%
\pgfpathrectangle{\pgfqpoint{0.495374in}{0.456793in}}{\pgfqpoint{4.393511in}{2.201537in}}%
\pgfusepath{clip}%
\pgfsetbuttcap%
\pgfsetroundjoin%
\pgfsetlinewidth{0.803000pt}%
\definecolor{currentstroke}{rgb}{0.690196,0.690196,0.690196}%
\pgfsetstrokecolor{currentstroke}%
\pgfsetdash{{0.800000pt}{1.320000pt}}{0.000000pt}%
\pgfpathmoveto{\pgfqpoint{0.495374in}{2.558260in}}%
\pgfpathlineto{\pgfqpoint{4.888885in}{2.558260in}}%
\pgfusepath{stroke}%
\end{pgfscope}%
\begin{pgfscope}%
\pgfsetbuttcap%
\pgfsetroundjoin%
\definecolor{currentfill}{rgb}{0.000000,0.000000,0.000000}%
\pgfsetfillcolor{currentfill}%
\pgfsetlinewidth{0.803000pt}%
\definecolor{currentstroke}{rgb}{0.000000,0.000000,0.000000}%
\pgfsetstrokecolor{currentstroke}%
\pgfsetdash{}{0pt}%
\pgfsys@defobject{currentmarker}{\pgfqpoint{-0.048611in}{0.000000in}}{\pgfqpoint{-0.000000in}{0.000000in}}{%
\pgfpathmoveto{\pgfqpoint{-0.000000in}{0.000000in}}%
\pgfpathlineto{\pgfqpoint{-0.048611in}{0.000000in}}%
\pgfusepath{stroke,fill}%
}%
\begin{pgfscope}%
\pgfsys@transformshift{0.495374in}{2.558260in}%
\pgfsys@useobject{currentmarker}{}%
\end{pgfscope}%
\end{pgfscope}%
\begin{pgfscope}%
\definecolor{textcolor}{rgb}{0.000000,0.000000,0.000000}%
\pgfsetstrokecolor{textcolor}%
\pgfsetfillcolor{textcolor}%
\pgftext[x=0.220682in, y=2.510035in, left, base]{\color{textcolor}\rmfamily\fontsize{10.000000}{12.000000}\selectfont \(\displaystyle {1.0}\)}%
\end{pgfscope}%
\begin{pgfscope}%
\definecolor{textcolor}{rgb}{0.000000,0.000000,0.000000}%
\pgfsetstrokecolor{textcolor}%
\pgfsetfillcolor{textcolor}%
\pgftext[x=0.165127in,y=1.557562in,,bottom,rotate=90.000000]{\color{textcolor}\rmfamily\fontsize{10.000000}{12.000000}\selectfont ..}%
\end{pgfscope}%
\begin{pgfscope}%
\pgfpathrectangle{\pgfqpoint{0.495374in}{0.456793in}}{\pgfqpoint{4.393511in}{2.201537in}}%
\pgfusepath{clip}%
\pgfsetbuttcap%
\pgfsetroundjoin%
\pgfsetlinewidth{1.505625pt}%
\definecolor{currentstroke}{rgb}{0.400000,0.760784,0.647059}%
\pgfsetstrokecolor{currentstroke}%
\pgfsetdash{{5.550000pt}{2.400000pt}}{0.000000pt}%
\pgfpathmoveto{\pgfqpoint{0.504161in}{0.556863in}}%
\pgfpathlineto{\pgfqpoint{2.349436in}{0.557724in}}%
\pgfpathlineto{\pgfqpoint{2.446093in}{0.559825in}}%
\pgfpathlineto{\pgfqpoint{2.639408in}{0.561486in}}%
\pgfpathlineto{\pgfqpoint{2.648195in}{0.564188in}}%
\pgfpathlineto{\pgfqpoint{2.656982in}{0.574616in}}%
\pgfpathlineto{\pgfqpoint{2.665769in}{0.597472in}}%
\pgfpathlineto{\pgfqpoint{2.674556in}{0.641263in}}%
\pgfpathlineto{\pgfqpoint{2.683343in}{0.710652in}}%
\pgfpathlineto{\pgfqpoint{2.692130in}{0.807040in}}%
\pgfpathlineto{\pgfqpoint{2.700917in}{0.929247in}}%
\pgfpathlineto{\pgfqpoint{2.762426in}{1.922890in}}%
\pgfpathlineto{\pgfqpoint{2.771213in}{2.036891in}}%
\pgfpathlineto{\pgfqpoint{2.780000in}{2.134380in}}%
\pgfpathlineto{\pgfqpoint{2.788787in}{2.216778in}}%
\pgfpathlineto{\pgfqpoint{2.797574in}{2.282805in}}%
\pgfpathlineto{\pgfqpoint{2.806361in}{2.330379in}}%
\pgfpathlineto{\pgfqpoint{2.815148in}{2.363322in}}%
\pgfpathlineto{\pgfqpoint{2.823935in}{2.386559in}}%
\pgfpathlineto{\pgfqpoint{2.832722in}{2.404191in}}%
\pgfpathlineto{\pgfqpoint{2.841509in}{2.416780in}}%
\pgfpathlineto{\pgfqpoint{2.859083in}{2.434873in}}%
\pgfpathlineto{\pgfqpoint{2.876657in}{2.446701in}}%
\pgfpathlineto{\pgfqpoint{2.894231in}{2.453686in}}%
\pgfpathlineto{\pgfqpoint{2.911805in}{2.456888in}}%
\pgfpathlineto{\pgfqpoint{2.938166in}{2.458990in}}%
\pgfpathlineto{\pgfqpoint{2.999676in}{2.460751in}}%
\pgfpathlineto{\pgfqpoint{3.166629in}{2.465555in}}%
\pgfpathlineto{\pgfqpoint{3.254499in}{2.477843in}}%
\pgfpathlineto{\pgfqpoint{3.324795in}{2.490152in}}%
\pgfpathlineto{\pgfqpoint{3.359943in}{2.499679in}}%
\pgfpathlineto{\pgfqpoint{3.412666in}{2.516230in}}%
\pgfpathlineto{\pgfqpoint{3.439027in}{2.522575in}}%
\pgfpathlineto{\pgfqpoint{3.465388in}{2.526818in}}%
\pgfpathlineto{\pgfqpoint{3.518110in}{2.530681in}}%
\pgfpathlineto{\pgfqpoint{3.632341in}{2.533222in}}%
\pgfpathlineto{\pgfqpoint{3.658702in}{2.536445in}}%
\pgfpathlineto{\pgfqpoint{3.728998in}{2.548934in}}%
\pgfpathlineto{\pgfqpoint{3.755359in}{2.551255in}}%
\pgfpathlineto{\pgfqpoint{3.799295in}{2.552556in}}%
\pgfpathlineto{\pgfqpoint{3.948674in}{2.554257in}}%
\pgfpathlineto{\pgfqpoint{4.177137in}{2.556859in}}%
\pgfpathlineto{\pgfqpoint{4.247433in}{2.557940in}}%
\pgfpathlineto{\pgfqpoint{4.449534in}{2.558220in}}%
\pgfpathlineto{\pgfqpoint{4.898885in}{2.558220in}}%
\pgfpathlineto{\pgfqpoint{4.898885in}{2.558220in}}%
\pgfusepath{stroke}%
\end{pgfscope}%
\begin{pgfscope}%
\pgfpathrectangle{\pgfqpoint{0.495374in}{0.456793in}}{\pgfqpoint{4.393511in}{2.201537in}}%
\pgfusepath{clip}%
\pgfsetrectcap%
\pgfsetroundjoin%
\pgfsetlinewidth{1.505625pt}%
\definecolor{currentstroke}{rgb}{0.988235,0.552941,0.384314}%
\pgfsetstrokecolor{currentstroke}%
\pgfsetdash{}{0pt}%
\pgfpathmoveto{\pgfqpoint{0.504161in}{0.556863in}}%
\pgfpathlineto{\pgfqpoint{2.656982in}{0.557063in}}%
\pgfpathlineto{\pgfqpoint{2.665769in}{0.558685in}}%
\pgfpathlineto{\pgfqpoint{2.674556in}{0.567954in}}%
\pgfpathlineto{\pgfqpoint{2.683343in}{0.597565in}}%
\pgfpathlineto{\pgfqpoint{2.692130in}{0.660128in}}%
\pgfpathlineto{\pgfqpoint{2.700917in}{0.756907in}}%
\pgfpathlineto{\pgfqpoint{2.718491in}{1.023778in}}%
\pgfpathlineto{\pgfqpoint{2.762426in}{1.749617in}}%
\pgfpathlineto{\pgfqpoint{2.780000in}{2.022295in}}%
\pgfpathlineto{\pgfqpoint{2.788787in}{2.142257in}}%
\pgfpathlineto{\pgfqpoint{2.797574in}{2.237374in}}%
\pgfpathlineto{\pgfqpoint{2.806361in}{2.308066in}}%
\pgfpathlineto{\pgfqpoint{2.815148in}{2.364203in}}%
\pgfpathlineto{\pgfqpoint{2.823935in}{2.403242in}}%
\pgfpathlineto{\pgfqpoint{2.832722in}{2.433914in}}%
\pgfpathlineto{\pgfqpoint{2.841509in}{2.458419in}}%
\pgfpathlineto{\pgfqpoint{2.850296in}{2.477358in}}%
\pgfpathlineto{\pgfqpoint{2.859083in}{2.492453in}}%
\pgfpathlineto{\pgfqpoint{2.867870in}{2.504485in}}%
\pgfpathlineto{\pgfqpoint{2.876657in}{2.513475in}}%
\pgfpathlineto{\pgfqpoint{2.885444in}{2.519621in}}%
\pgfpathlineto{\pgfqpoint{2.894231in}{2.524366in}}%
\pgfpathlineto{\pgfqpoint{2.911805in}{2.530192in}}%
\pgfpathlineto{\pgfqpoint{2.938166in}{2.535197in}}%
\pgfpathlineto{\pgfqpoint{2.964527in}{2.538740in}}%
\pgfpathlineto{\pgfqpoint{3.061185in}{2.548210in}}%
\pgfpathlineto{\pgfqpoint{3.131481in}{2.551633in}}%
\pgfpathlineto{\pgfqpoint{3.263286in}{2.553836in}}%
\pgfpathlineto{\pgfqpoint{3.562045in}{2.555517in}}%
\pgfpathlineto{\pgfqpoint{3.799295in}{2.556598in}}%
\pgfpathlineto{\pgfqpoint{3.931100in}{2.556999in}}%
\pgfpathlineto{\pgfqpoint{4.396812in}{2.558020in}}%
\pgfpathlineto{\pgfqpoint{4.898885in}{2.558043in}}%
\pgfpathlineto{\pgfqpoint{4.898885in}{2.558043in}}%
\pgfusepath{stroke}%
\end{pgfscope}%
\begin{pgfscope}%
\pgfpathrectangle{\pgfqpoint{0.495374in}{0.456793in}}{\pgfqpoint{4.393511in}{2.201537in}}%
\pgfusepath{clip}%
\pgfsetbuttcap%
\pgfsetroundjoin%
\pgfsetlinewidth{1.505625pt}%
\definecolor{currentstroke}{rgb}{0.552941,0.627451,0.796078}%
\pgfsetstrokecolor{currentstroke}%
\pgfsetdash{{9.600000pt}{2.400000pt}{1.500000pt}{2.400000pt}}{0.000000pt}%
\pgfpathmoveto{\pgfqpoint{0.504161in}{0.556863in}}%
\pgfpathlineto{\pgfqpoint{2.920592in}{0.557324in}}%
\pgfpathlineto{\pgfqpoint{2.929379in}{0.558384in}}%
\pgfpathlineto{\pgfqpoint{2.938166in}{0.563468in}}%
\pgfpathlineto{\pgfqpoint{2.946953in}{0.575877in}}%
\pgfpathlineto{\pgfqpoint{2.955740in}{0.597452in}}%
\pgfpathlineto{\pgfqpoint{2.964527in}{0.635579in}}%
\pgfpathlineto{\pgfqpoint{2.973314in}{0.697683in}}%
\pgfpathlineto{\pgfqpoint{2.982101in}{0.783163in}}%
\pgfpathlineto{\pgfqpoint{2.990889in}{0.890439in}}%
\pgfpathlineto{\pgfqpoint{2.999676in}{1.011525in}}%
\pgfpathlineto{\pgfqpoint{3.017250in}{1.295186in}}%
\pgfpathlineto{\pgfqpoint{3.052398in}{1.862367in}}%
\pgfpathlineto{\pgfqpoint{3.061185in}{1.989077in}}%
\pgfpathlineto{\pgfqpoint{3.078759in}{2.194803in}}%
\pgfpathlineto{\pgfqpoint{3.087546in}{2.274439in}}%
\pgfpathlineto{\pgfqpoint{3.096333in}{2.338945in}}%
\pgfpathlineto{\pgfqpoint{3.105120in}{2.387860in}}%
\pgfpathlineto{\pgfqpoint{3.113907in}{2.424626in}}%
\pgfpathlineto{\pgfqpoint{3.122694in}{2.451525in}}%
\pgfpathlineto{\pgfqpoint{3.131481in}{2.471559in}}%
\pgfpathlineto{\pgfqpoint{3.149055in}{2.502281in}}%
\pgfpathlineto{\pgfqpoint{3.157842in}{2.513088in}}%
\pgfpathlineto{\pgfqpoint{3.166629in}{2.521394in}}%
\pgfpathlineto{\pgfqpoint{3.184203in}{2.533483in}}%
\pgfpathlineto{\pgfqpoint{3.201777in}{2.540247in}}%
\pgfpathlineto{\pgfqpoint{3.228138in}{2.544771in}}%
\pgfpathlineto{\pgfqpoint{3.307221in}{2.550655in}}%
\pgfpathlineto{\pgfqpoint{3.386305in}{2.554137in}}%
\pgfpathlineto{\pgfqpoint{3.579619in}{2.555718in}}%
\pgfpathlineto{\pgfqpoint{3.878378in}{2.556679in}}%
\pgfpathlineto{\pgfqpoint{4.898885in}{2.558260in}}%
\pgfpathlineto{\pgfqpoint{4.898885in}{2.558260in}}%
\pgfusepath{stroke}%
\end{pgfscope}%
\begin{pgfscope}%
\pgfsetrectcap%
\pgfsetmiterjoin%
\pgfsetlinewidth{0.803000pt}%
\definecolor{currentstroke}{rgb}{0.000000,0.000000,0.000000}%
\pgfsetstrokecolor{currentstroke}%
\pgfsetdash{}{0pt}%
\pgfpathmoveto{\pgfqpoint{0.495374in}{0.456793in}}%
\pgfpathlineto{\pgfqpoint{0.495374in}{2.658330in}}%
\pgfusepath{stroke}%
\end{pgfscope}%
\begin{pgfscope}%
\pgfsetrectcap%
\pgfsetmiterjoin%
\pgfsetlinewidth{0.803000pt}%
\definecolor{currentstroke}{rgb}{0.000000,0.000000,0.000000}%
\pgfsetstrokecolor{currentstroke}%
\pgfsetdash{}{0pt}%
\pgfpathmoveto{\pgfqpoint{4.888885in}{0.456793in}}%
\pgfpathlineto{\pgfqpoint{4.888885in}{2.658330in}}%
\pgfusepath{stroke}%
\end{pgfscope}%
\begin{pgfscope}%
\pgfsetrectcap%
\pgfsetmiterjoin%
\pgfsetlinewidth{0.803000pt}%
\definecolor{currentstroke}{rgb}{0.000000,0.000000,0.000000}%
\pgfsetstrokecolor{currentstroke}%
\pgfsetdash{}{0pt}%
\pgfpathmoveto{\pgfqpoint{0.495374in}{0.456793in}}%
\pgfpathlineto{\pgfqpoint{4.888885in}{0.456793in}}%
\pgfusepath{stroke}%
\end{pgfscope}%
\begin{pgfscope}%
\pgfsetrectcap%
\pgfsetmiterjoin%
\pgfsetlinewidth{0.803000pt}%
\definecolor{currentstroke}{rgb}{0.000000,0.000000,0.000000}%
\pgfsetstrokecolor{currentstroke}%
\pgfsetdash{}{0pt}%
\pgfpathmoveto{\pgfqpoint{0.495374in}{2.658330in}}%
\pgfpathlineto{\pgfqpoint{4.888885in}{2.658330in}}%
\pgfusepath{stroke}%
\end{pgfscope}%
\begin{pgfscope}%
\pgfsetbuttcap%
\pgfsetmiterjoin%
\definecolor{currentfill}{rgb}{1.000000,1.000000,1.000000}%
\pgfsetfillcolor{currentfill}%
\pgfsetfillopacity{0.800000}%
\pgfsetlinewidth{1.003750pt}%
\definecolor{currentstroke}{rgb}{0.800000,0.800000,0.800000}%
\pgfsetstrokecolor{currentstroke}%
\pgfsetstrokeopacity{0.800000}%
\pgfsetdash{}{0pt}%
\pgfpathmoveto{\pgfqpoint{0.564819in}{1.966201in}}%
\pgfpathlineto{\pgfqpoint{1.734573in}{1.966201in}}%
\pgfpathlineto{\pgfqpoint{1.734573in}{2.588886in}}%
\pgfpathlineto{\pgfqpoint{0.564819in}{2.588886in}}%
\pgfpathlineto{\pgfqpoint{0.564819in}{1.966201in}}%
\pgfpathclose%
\pgfusepath{stroke,fill}%
\end{pgfscope}%
\begin{pgfscope}%
\pgfsetbuttcap%
\pgfsetroundjoin%
\pgfsetlinewidth{1.505625pt}%
\definecolor{currentstroke}{rgb}{0.400000,0.760784,0.647059}%
\pgfsetstrokecolor{currentstroke}%
\pgfsetdash{{5.550000pt}{2.400000pt}}{0.000000pt}%
\pgfpathmoveto{\pgfqpoint{0.620374in}{2.484719in}}%
\pgfpathlineto{\pgfqpoint{0.759263in}{2.484719in}}%
\pgfpathlineto{\pgfqpoint{0.898152in}{2.484719in}}%
\pgfusepath{stroke}%
\end{pgfscope}%
\begin{pgfscope}%
\definecolor{textcolor}{rgb}{0.000000,0.000000,0.000000}%
\pgfsetstrokecolor{textcolor}%
\pgfsetfillcolor{textcolor}%
\pgftext[x=1.009263in,y=2.436108in,left,base]{\color{textcolor}\rmfamily\fontsize{10.000000}{12.000000}\selectfont Bare-metal}%
\end{pgfscope}%
\begin{pgfscope}%
\pgfsetrectcap%
\pgfsetroundjoin%
\pgfsetlinewidth{1.505625pt}%
\definecolor{currentstroke}{rgb}{0.988235,0.552941,0.384314}%
\pgfsetstrokecolor{currentstroke}%
\pgfsetdash{}{0pt}%
\pgfpathmoveto{\pgfqpoint{0.620374in}{2.291046in}}%
\pgfpathlineto{\pgfqpoint{0.759263in}{2.291046in}}%
\pgfpathlineto{\pgfqpoint{0.898152in}{2.291046in}}%
\pgfusepath{stroke}%
\end{pgfscope}%
\begin{pgfscope}%
\definecolor{textcolor}{rgb}{0.000000,0.000000,0.000000}%
\pgfsetstrokecolor{textcolor}%
\pgfsetfillcolor{textcolor}%
\pgftext[x=1.009263in,y=2.242435in,left,base]{\color{textcolor}\rmfamily\fontsize{10.000000}{12.000000}\selectfont TSN-CNI}%
\end{pgfscope}%
\begin{pgfscope}%
\pgfsetbuttcap%
\pgfsetroundjoin%
\pgfsetlinewidth{1.505625pt}%
\definecolor{currentstroke}{rgb}{0.552941,0.627451,0.796078}%
\pgfsetstrokecolor{currentstroke}%
\pgfsetdash{{9.600000pt}{2.400000pt}{1.500000pt}{2.400000pt}}{0.000000pt}%
\pgfpathmoveto{\pgfqpoint{0.620374in}{2.097373in}}%
\pgfpathlineto{\pgfqpoint{0.759263in}{2.097373in}}%
\pgfpathlineto{\pgfqpoint{0.898152in}{2.097373in}}%
\pgfusepath{stroke}%
\end{pgfscope}%
\begin{pgfscope}%
\definecolor{textcolor}{rgb}{0.000000,0.000000,0.000000}%
\pgfsetstrokecolor{textcolor}%
\pgfsetfillcolor{textcolor}%
\pgftext[x=1.009263in,y=2.048762in,left,base]{\color{textcolor}\rmfamily\fontsize{10.000000}{12.000000}\selectfont Flannel}%
\end{pgfscope}%
\end{pgfpicture}%
\makeatother%
\endgroup%